\documentclass{cpbtex}

\usepackage{graphicx}% Include figure files
\usepackage{subcaption}
\usepackage{caption}
\usepackage{dcolumn}% Align table columns on decimal point
\usepackage{bm}% bold math
\usepackage[inline]{enumitem}

\newcommand{\myeqref}[1]{Eq.~\eqref{#1}}
\newcommand{\mySecRef}[1]{Sec.~\ref{#1}~~~}
\newcommand{\mycite}[1]{\ucite{#1}}
\newcommand{\myRefInd}[1]{Ref.~\cite{#1}}
\newcommand{\rvs}[1] { #1}
\newcommand{\mrvs}[1] { #1}
\begin{document}

\title{Derivation of lattice Boltzmann equation via analytical \\ characteristic integral\thanks{Project supported by National Science and Technology Major Project (Grant No.~2017ZX06002002).}}% Force line breaks with \\

	\author{Huanfeng Ye$^{1}$\thanks{Corresponding author. E-mail: huanfye@163.com}, \ Bo Kuang$^{1}$ \ and \ Yanhua Yang$^{1,2}$\\
	$^{1}${School of Nuclear Science and Engineering, }\\
	{Shanghai Jiao Tong University, Shanghai 200240, China}\\  % The line break was forced via \\
	$^{2}${National Energy Key Laboratory of Nuclear Power Software,}\\
	{ Beijing 102209, China}}

\date{\today}% It is always \today, today,
\maketitle

\begin{abstract}
	A lattice Boltzmann (LB) theory, analytical characteristic integral (ACI) LB theory, is proposed in this paper. ACI LB theory takes Bhatnagar-Gross-Krook (BGK) Boltzmann equation as the exact kinetic equation behind Navier-Stokes continuum and momentum equations and  constructs LB equation by rigorously integrating BGK-Boltzmann equation along characteristics. It's a general theory, supporting most existed LB equations including the standard lattice BGK (LBGK) equation inherited from lattice-gas automata, whose theoretical foundation had been questioned. ACI LB theory also indicates that the characteristic parameter of LB equation is collision number, depicting the particle-interacting intensity in the time span of LB equation, instead of traditionally assumed relaxation time, and the over relaxation time problem is merely a manifestation of temporal evolution of equilibrium distribution along characteristics under high collision number, irrelevant to particle kinetics. In ACI LB theory, the temporal evolution of equilibrium distribution along characteristics is the determinant of LB method accuracy and we numerically prove it. 
\end{abstract}

%\pacs{23.23.+x, 56.65.Dy}% PACS, the Physics and Astronomy



%\tableofcontents

\section{Introduction}
\label{sec:intro}
In the past few decades, the lattice Boltzmann (LB) method  has been proven as an efficient alternative approach \mrvs{to} many computational fluid dynamics (CFD) areas, e.g. hydrodynamic systems\mycite{LiuZou-178,BaochangZhaoli-179}, multiphase and multicomponent fluids\mycite{LaddVerberg-186,SheikholeslamiGorji-Bandpy-163,LiuCheng-181,SemmaElGanaoui-240,MountrakisLorenz-483}, porous media flow\mycite{MengGuo-2046,LiuHe-2047}. With the rapid development of multi-core super computer and parallel computation, LB method is gaining more attention from CFD researchers due to its inherent parallelism\mycite{ZhangSun-2775,WuLiu-2776,SunJiang-2777}. 
Historically, LB method arose from the pioneering idea of lattice-gas automata\mycite{FrischHasslacher-2751}, however more recently we tend to link it with the kinetic equation, Bhatnagar-Gross-Krook (BGK) Boltzmann equation\mycite{BhatnagarGross-95}, which leads to the reconstruction of LB theory\mycite{Dellar-2074,HeChen-173,Hwang-2023,YongZhao-2019,BoeschKarlin-2075,HeLuo-2016,Shan-2582}.

\mrvs{Although} the newly constructed LB theory has abandoned the original lattice-gas theory, the interest to reproduce the classical lattice BGK (LBGK) equation inherited from lattice-gas automata, which has been proven numerically stable, is still considerable. To achieve this, numerous approaches \mrvs{have} been proposed, e.g. Strang splitting\mycite{Dellar-2074}, Maxwell iteration\mycite{YongZhao-2019}, Taylor expansion\mycite{HeLuo-2016, SterlingChen-252}, etc. 
However, the approximation and truncation employed in these approaches make the reproduction unconvincing.
\rvs{Taking the most popular Taylor expansion scheme as an instance, LB equation is dealt as a characteristic integral of BGK Boltzmann equation with constant collision term. To fix the truncation error introduced by constant collision term assumption, the scheme expands the left-hand-side integral of BGK Boltzmann equation in Taylor series during recovering Navier-Stokes equations with Chapmann-Enskog expansion\mycite{SterlingChen-252}. Employing the 2nd-degree Taylor expansion, the LBGK equation can be properly recovered.  
But the truncated Taylor series introduce uncertainty.
And a further research\mycite{HeChen-173} asserts that employing Taylor expansion scheme to recover Navier-Stokes energy equation, the viscosity in viscous heat dissipation term is inconsistent with momentum equation.  
Hence {\it He et al}\mycite{HeChen-173} employ a trapezoidal rule to improve the approximate accuracy of collision term to avoid Taylor expansion. Recently, {\it Boesch and Karlin}\mycite{BoeschKarlin-2075} use Euler-Maclaurin integral to analytically integrate the collision term, eliminating the truncation error of trapezoidal rule. Compared with trapezoidal rule, the result of Euler-Maclaurin integral ({\it exact} LB equation called in the paper),  is theoretically more rigorous. 
The problem is that the elegant but extremely complicated Euler-Maclaurin integral cloaks the important effect of equilibrium distribution. It makes the derivation in  \myRefInd{BoeschKarlin-2075} be founded on a quite rough approximation of equilibrium distribution, which leads {\it Boesch and Karlin} to the conclusion that classical LBGK equation can not be recovered from analytical characteristic integral of  BGK-Boltzman equation. Especially, when the viscosity is minimal, the relaxation time in LBGK equation would be greater than 1 tending to 2,  i.e. overrelaxation, meanwhile it is always under 1 in {\it exact} LB equation. Thus {\it Boesch and Karlin}\mycite{BoeschKarlin-2075} assert that the overrelaxation in  LBGK equation is beyond the kinetic theory of continuous-time  BGK-Boltzmann equation.}  
Then an essential question arises: can LBGK equation be exactly reproduced based on BGK-Boltzmann equation?

In this paper, we propose \mrvs{an} LB theory to link the LB method and BGK-Boltzmann equation, \rvs{designated as analytical characteristics integral (ACI) LB theory.
ACI theory follows the philosophy of \myRefInd{BoeschKarlin-2075, HeChen-173} to construct LB equation, i.e. improving the accuracy of collision term integral to eliminate the truncation error and inconsistence in Taylor expansion scheme. 
Through analyzing BGK-Boltzmann equation, ACI theory asserts that BGK-Boltzmann equation is accurate enough to describe the particle kinetics behind Navier-Stokes continuum and momentum equations, and LB equation can be constructed via analytically integrating BGK-Boltzmann equation along characteristics with approximated temporal evolution of equilibrium distribution.  
ACI theory employs  differential equation solving skill to achieve the analytical integral. 
Compared with Euler-Maclaurin integral employed in \myRefInd{BoeschKarlin-2075}, differential equation solving skill is mathematically more concise with a clear physical figure. 
Meanwhile the rigorous characteristic integral avoids the truncation error of trapezoidal rule in \myRefInd{HeChen-173}.
In ACI theory, the temporal evolution of equilibrium distribution along characteristics is the kernel of constructing LB equation, determining the computational stability and accuracy of LB equations. It extends the generality of BGK-Boltzmann kinetic theory discussed in \myRefInd{BoeschKarlin-2075}.
To demonstrate it, we recover several popular LB equations and numerically analyze them. }
Our derivation also indicates that the characteristic parameter of LB equation is collision number instead of the traditionally assumed relaxation time, which is merely a reflection of the temporal evolution of equilibrium distribution along characteristics, and the over relaxation time problem is a manifestation of the evolution model behind LBGK equation under high collision number, \mrvs{supported by the kinetic theory of continuous-time  BGK-Boltzmann equation.}

This paper is organized as follows. In \mySecRef{sec:BGK}, we propose ACI LB theory and recover  popular LB equations. In \mySecRef{sec:disCrt},  we demonstrate the discretion of LB equation under ACI LB theory. In \mySecRef{sec:numRe}, we numerically analyze the derived LB equations with several simple benchmarks. \mySecRef{sec:con} concludes the paper.

\section{Analytical characteristic integral LB theory}
\label{sec:BGK}
BGK-Boltzmann equation is a result of linear approximation on collision term \mrvs{of}  Boltzmann equation\mycite{BhatnagarGross-95},
\begin{equation}\label{eq:bgk}
% MathType!Translator!2!1!LaTeX.tdl!LaTeX 2.09 and later!
\frac{{\partial f}}{{\partial t}} + \vec \xi  \cdot \frac{{\partial f}}{{\partial \vec r}} =  - \frac{1}{\lambda }\left( {f - g} \right),% MathType!End!2!1!
\end{equation}
where % MathType!Translator!2!1!LaTeX.tdl!LaTeX 2.09 and later!
$f \equiv f\left( {\vec r,\vec \xi ,t} \right)$ % MathType!End!2!1!
is the particle distribution, % MathType!Translator!2!1!LaTeX.tdl!LaTeX 2.09 and later!
$\vec \xi $ % MathType!End!2!1!
is particle microscopic velocity, % MathType!Translator!2!1!LaTeX.tdl!LaTeX 2.09 and later!
$\lambda $ % MathType!End!2!1!
is the collision time, and $g$
is the Maxwell-Boltzmann distribution,
\begin{equation}\label{eq:maxw}
%MathType!Translator!2!1!LaTeX.tdl!LaTeX 2.09 and later!
g \equiv \frac{\rho }{{{{\left( {2\pi RT} \right)}^{D/2}}}}\exp \left( { - \frac{{{{\left( {\vec \xi  - \vec u} \right)}^2}}}{{2RT}}} \right),% MathType!End!2!1!
\end{equation}
where $R$ is the ideal gas constant, $D$ is the dimension of the space, and $\rho $, $\vec u$, and $T$ are the macroscopic density of mass, velocity and temperature, respectively. The macroscopic variables, $\rho $, $\vec u$, $T$ are the (microscopic velocity) moments of the distribution function $f$:
\begin{subequations}\label{eq:macro}
	\begin{equation}\label{eq:rho}
	% MathType!Translator!2!1!LaTeX.tdl!LaTeX 2.09 and later!
	\rho  = \int {fd\vec \xi }  = \int {gd\vec \xi },% MathType!End!2!1!
	\end{equation}
	\begin{equation}\label{eq:vel}
	% MathType!Translator!2!1!LaTeX.tdl!LaTeX 2.09 and later!
	\rho \vec u = \int {\vec \xi fd\vec \xi }  = \int {\vec \xi gd\vec \xi },% MathType!End!2!1!
	\end{equation}
	\begin{equation}\label{eq:temp}
	% MathType!Translator!2!1!LaTeX.tdl!LaTeX 2.09 and later!
	\rho RT = \frac{1}{2}\int {{{\left( {\vec \xi  - \vec u} \right)}^2}fd\vec \xi }  = \frac{1}{2}\int {{{\left( {\vec \xi  - \vec u} \right)}^2}gd\vec \xi }.% MathType!End!2!1!
	\end{equation}
\end{subequations}

BGK-Boltzmann equation is a kinetic equation describing the transient interaction between particles. In LB theory, we use it to depict the particle kinetics behind Navier-Stokes continuum and momentum equations. It can be mathematically validated with classical Chapman-Enskog (CE) expansion\mycite{ChapmanCowling-93}.
CE expansion is a multiple scale technique, proposed to solve the Boltzmann equation at first\mycite{ChapmanCowling-93}. The key concept of CE expansion is formulating the flow phenomena with different time scales. It decomposes the partial \mrvs{derivatives} with respect to $t$ and ${\vec r}$ into different scales,
\begin{subequations}\label{eq:deScl}
	\begin{equation}\label{eq:timScl}
	% MathType!Translator!2!1!LaTeX.tdl!LaTeX 2.09 and later!
	\frac{\partial }{{\partial t}} = \varepsilon \frac{\partial }{{\partial {t_1}}} + {\varepsilon ^2}\frac{\partial }{{\partial {t_2}}} + ... ~,% MathType!End!2!1!
	\end{equation}
	\begin{equation}\label{eq:spcScl}
	% MathType!Translator!2!1!LaTeX.tdl!LaTeX 2.09 and later!
	\frac{\partial }{{\partial \vec r}} = \varepsilon \frac{\partial }{{\partial {{\vec r}_1}}},% MathType!End!2!1!
	\end{equation}
\end{subequations}
where $\varepsilon $ is a parameter who is small enough to identify the scales, $t_i$ and ${\vec r}_1$ \mrvs{are} the relative time and space scales respectively. Specifically, $t_1$ and $t_2$ represent the time scales of convection and diffusion respectively\mycite{Hwang-2023}. Similarly, we expand the particle distribution $f$ with parameter $\varepsilon $,
\begin{equation}\label{eq:fScl}
% MathType!Translator!2!1!LaTeX.tdl!LaTeX 2.09 and later!
f = g + \sum\limits_{n = 1}^\infty  {{\varepsilon ^n}{f^{\left( n \right)}}},% MathType!End!2!1!
\end{equation}
where $g$ is the Maxwell-Boltzmann distribution. The expansion implies that the deviation between particle distribution $f$ and equilibrium distribution $g$ is limited. Combining with the macroscopic variable formulas in \myeqref{eq:macro}, the integrals of the expansion for macroscopic variables yield
\begin{subequations}\label{eq:NEqMom}
	\begin{equation}\label{eq:NEqRho}
	% MathType!Translator!2!1!LaTeX.tdl!LaTeX 2.09 and later!
	\int {\sum\limits_{n = 1}^\infty  {{\varepsilon ^n}{f^{\left( n \right)}}} d\vec \xi }  = 0,% MathType!End!2!1!
	\end{equation}
	\begin{equation}\label{eq:NEqVel}
	% MathType!Translator!2!1!LaTeX.tdl!LaTeX 2.09 and later!
	\int {\sum\limits_{n = 1}^\infty  {{\varepsilon ^n}{f^{\left( n \right)}}} } \vec \xi d\vec \xi  = \vec 0,% MathType!End!2!1!
	\end{equation}
	\begin{equation}\label{eq:NEqTemp}
	% MathType!Translator!2!1!LaTeX.tdl!LaTeX 2.09 and later!
	\int {\sum\limits_{n = 1}^\infty  {{\varepsilon ^n}{f^{\left( n \right)}}} {{\left( {\vec \xi  - \vec u} \right)}^2}d\vec \xi }  = 0.% MathType!End!2!1!
	\end{equation}
\end{subequations}
In order to solve the BGK-Boltzmann equation, CE expansion employs an extra assumption that the first three velocity moment integrals of the decomposed distribution equal zero on each order of $\varepsilon $. For sake of simplicity but without losing generality, we use a linear combination of first three moments $\phi \left( {\vec \xi } \right)$ to describe the assumption,
\begin{equation}\label{eq:MomAssp}
% MathType!Translator!2!1!LaTeX.tdl!LaTeX 2.09 and later!
\int {{f^{\left( n \right)}}} \phi \left( {\vec \xi } \right)d\vec \xi  = 0,% MathType!End!2!1!
\end{equation}
where
% MathType!Translator!2!1!LaTeX.tdl!LaTeX 2.09 and later!
\[\phi \left( {\vec \xi } \right) = A + \vec B \cdot \vec \xi  + C{\xi ^2},\]% MathType!End!2!1!
where $A$ and $C$ are arbitrary constants, $\vec B$ is arbitrary constant vector.

Now, \mrvs{we substitute} the expansions in \myeqref{eq:deScl} and \myeqref{eq:fScl} into BGK-Boltzmann equation in \myeqref{eq:bgk} and \mrvs{collect} the terms with same order of $\varepsilon $ to generate  new equations on difference scales. The first two scaling equations of $\varepsilon $ read:
\begin{subequations}\label{eq:CEBGK}
	\begin{equation}\label{eq:CEBGK1}
	% MathType!Translator!2!1!LaTeX.tdl!LaTeX 2.09 and later!
	\frac{{\partial g}}{{\partial {t_1}}} + \vec \xi  \cdot \frac{{\partial g}}{{\partial {{\vec r}_1}}} =  - \frac{1}{\lambda }{f^{\left( 1 \right)}},% MathType!End!2!1!
	\end{equation}
	\begin{equation}\label{eq:CEBGK2}
	% MathType!Translator!2!1!LaTeX.tdl!LaTeX 2.09 and later!
	\frac{{\partial g}}{{\partial {t_2}}} + \frac{{\partial {f^{\left( 1 \right)}}}}{{\partial {t_1}}} + \vec \xi  \cdot \frac{{\partial {f^{\left( 1 \right)}}}}{{\partial {{\vec r}_1}}} =  - \frac{1}{\lambda }{f^{\left( 2 \right)}}.% MathType!End!2!1!
	\end{equation}
\end{subequations}
Integrating \myeqref{eq:CEBGK} on the first two velocity moments with the assumption in \myeqref{eq:MomAssp}, we can get
\begin{subequations}\label{eq:intCEBGK1}
	\begin{equation}\label{eq:intCEBGK1rho}
	\frac{{\partial \rho }}{{\partial {t_1}}} + \frac{{\partial \rho \vec u}}{{\partial {{\vec r}_1}}} = 0,
	\end{equation}
	\begin{equation}\label{eq:intCEBGK1Vel}
	\frac{{\partial \rho \vec u}}{{\partial {t_1}}} + \frac{{\partial \rho \vec u\vec u}}{{\partial {{\vec r}_1}}} + \frac{{\partial \rho RT}}{{\partial {{\vec r}_1}}} = 0;
	\end{equation}
\end{subequations}
\begin{subequations}\label{eq:intCEBGK2}
	\begin{equation}\label{eq:intCEBGK2rho}
	\frac{{\partial \rho }}{{\partial {t_2}}} = 0,
	\end{equation}
	\begin{equation}\label{eq:intCEBGK2Vel}
	\frac{{\partial \rho \vec u}}{{\partial {t_2}}} + \frac{\partial }{{\partial \vec r}} \cdot \left( { - \rho RT\lambda \left( {\left[ {\nabla \vec u} \right] + {{\left[ {\nabla \vec u} \right]}^T}} \right)} \right) = 0.
	\end{equation}
\end{subequations}
Incorporating the formulas in \myeqref{eq:intCEBGK1} and \myeqref{eq:intCEBGK2} based on their \mrvs{orders} of velocity moment respectively, the macroscopic hydrodynamic equations are recovered for BGK-Boltzmann equation with Maxwell-Boltzmann distribution,
\begin{subequations}\label{eq:NS}
	\begin{equation}\label{eq:NSconti}
	\frac{{\partial \rho }}{{\partial t}} + \frac{{\partial \rho \vec u}}{{\partial \vec r}} = 0,
	\end{equation}
	\begin{equation}\label{eq:NSVelo}
	\frac{{\partial \rho }}{{\partial t}} + \frac{{\partial \rho \vec u\vec u}}{{\partial \vec r}} = - \frac{{\partial p}}{{\partial \vec r}} 
	+ \frac{\partial }{{\partial \vec r}} \cdot \left( {\nu \rho \left( {\left[ {\nabla \vec u} \right] + {{\left[ {\nabla \vec u} \right]}^T}} \right)} \right),
	\end{equation}
\end{subequations}
where $p=\rho RT$ is the pressure, $\nu =RT\lambda $ is the kinematic viscosity. Then  the collision time $\lambda $ can be solved as
\begin{equation}\label{eq:clTm}
% MathType!Translator!2!1!LaTeX.tdl!LaTeX 2.09 and later!
\lambda  = \frac{\nu }{{RT}}.% MathType!End!2!1!
\end{equation}

\rvs{It should be noted that this recovery of Navier-Stokes equations did not involve Taylor expansion, avoiding the truncation error.} The derivation shows that BGK-Boltzmann equation is accurate enough to describe the particle kinetics behind Navier-Stokes continuum and momentum equations. Assuming BGK-Boltzmann equation is the exact kinetic equation for these two equations, all we need to do is constructing LB equation based on BGK-Boltzmann equation.
\mrvs{Here we follow the philosophy of Boesh-Karlin approach\mycite{BoeschKarlin-2075}, i.e. constructing LB equation through analytically integrating BGK-Boltzmann equation along characteristics. Actually, with simple equation transformation\mycite{HeLuo-2016},  all terms in BGK-Boltzmann equation can be directly integrated except the equilibrium term (Maxwell-Boltzmann distribution).} We start with transforming the left side of BGK-Boltzmann equation, treating it as the temporal derivative along characteristic line $\vec \xi $, 
\begin{equation}
\label{eq:ODE}
\frac{{df}}{{dt}} = - \frac{1}{\lambda}\left(f-g\right),
\end{equation}
where
% MathType!Translator!2!1!LaTeX.tdl!LaTeX 2.09 and later!
\[\frac{d}{{dt}} \equiv \frac{\partial }{{\partial t}} + \vec \xi  \cdot \frac{\partial }{{\partial \vec r}}.\]% MathType!End!2!1!
The \myeqref{eq:ODE} can be converted into integrable form by introducing an integrating factor $e^{t/\lambda}$,
\begin{equation}\label{eq:refomBGK}
% MathType!Translator!2!1!LaTeX.tdl!LaTeX 2.09 and later!
\frac{d}{{dt}}{e^{t/\lambda }}f = \frac{1}{\lambda }{e^{t/\lambda }}g.% MathType!End!2!1!
\end{equation}
Hence LB equation can be generated by directly integrating BGK-Boltzmann equation along characteristics, it reads
\begin{equation}\label{eq:BGKint}
{e^{\Delta t/\lambda }}{f^{n + 1}} - {f^n} = \int\limits_0^{\Delta t} {\frac{1}{\lambda }{e^{t'/\lambda }}g\left( {t'} \right)dt'}.
\end{equation}
It should be noted that we have introduced short notations to simplify the expression in \myeqref{eq:BGKint},
\begin{align*}
{f^{n + 1}} &= f\left( {\vec r + \vec \xi \Delta t,\vec \xi ,t + \Delta t} \right),  \\
{f^n} &= f\left( {\vec r,\vec \xi ,t} \right),  \\
g\left( {t'} \right) &= g\left( {\vec r + \vec \xi t',\vec \xi ,t + t'} \right).
\end{align*}
This notating convention would be kept in the following and extended into other variables. 

The integration in \myeqref{eq:BGKint} hasn't been finished, where the integral of $g\left( {t'} \right)$  still remains unsolved. 
Since the calculation of $g\left( {t'} \right)$ involves macroscopic density $\rho$ and velocity $\vec u$ (the temperature $T$ would be dealt as a field constant in practice), which \mrvs{evolute} nonlinearly and need to be calculated by LB equation, it is impossible to get its analytical distribution along the characteristics. The only solution is using approximation. Since BGK-Boltzmann equation is assumed exact for the particle kinetics behind Navier-Stokes continuum and momentum equation, and LB equation is its analytical integration along characteristics, then the accuracy of LB equation solely depends on the design of $g\left( {t'} \right)$ evolution along characteristics, i.e. the aforementioned temporal evolution of equilibrium distribution.
In other words $g\left( {t'} \right)$ model forges the final form of LB equation. 
\mrvs{It greatly extends the kinetic theory of BGK-Boltzmann equation discussed in Boesh-Karlin approach\mycite{BoeschKarlin-2075}, which we will detailedly demonstrate in the following. 
In case the evolution of BGK-Boltzmann equation departs the physical process too much, we restrict the value of $g\left( {t'} \right)$ at $t'=0$ to $g^n$ , ensuring that the transient equation of particle interaction with respect to $\left( {\vec r,\vec \xi ,t} \right)$ is exact BGK-Boltzmann equation.}
We call this LB theory as analytical characteristic integral (ACI) LB theory.

We start with typical dealing for an unknown distribution, i.e. assuming $g\left( {t'} \right)$  constant over $[0,\Delta t]$, namely steady assumption (SA) model,
\begin{equation}
g\left( {t'} \right) = {g^n}, t'\in\left[0,\Delta t\right],
\end{equation}
then its LB equation reads
\begin{equation}\label{eq:SALB}
{f^{n + 1}} - {f^n} =  - \left( {1 - {e^{ - \Delta t/\lambda }}} \right)\left( {{f^n} - {g^n}} \right).
\end{equation}
\rvs{\myRefInd{HeLuo-2016} proposes} linear interpolation between $g^n$ and $g^{n+1}$, 
\begin{equation}
g\left( {t'} \right) = {g^n} + \frac{{t'}}{{\Delta t}}\left( {{g^{n + 1}} - {g^n}} \right).
\end{equation}
Substituting it into \myeqref{eq:BGKint}, we can get its corresponding LB equation,
\begin{align}
\label{eq:CABKLBM}
{f^{n + 1}} - {f^n} = & - \frac{1}{2}\left( {\frac{{\Delta t}}{\lambda }} \right)\left( {\left( {{f^{n + 1}} - g^{n+1}} \right) + \left( {{f^n} - g^n} \right)} \right) \nonumber \\
& - \left( {\frac{{ - \frac{{\Delta t}}{\lambda }}}{{{e^{ - \frac{{\Delta t}}{\lambda }}} - 1}} - 1 - \frac{{\Delta t}}{{2\lambda }}} \right)\left( {{f^{n + 1}} - {f^n}} \right).
\end{align}
\rvs{It worths noting that though \myRefInd{HeLuo-2016} got this analytical integration, the authors had expanded it in Taylor series to recover LBGK equation form,
\begin{equation}
f^{n+1}-f^{n}=-\frac{\Delta t}{\lambda}\left(f^n-g^n\right)
\end{equation}
 which makes the analytical integral of collision term degrade into constant approximation. In order to fix the truncation error introduced by constant collision term approximation, Taylor expansion is still inevitable during recovery of Navier-Stokes equations. Therefore, the scheme in \myRefInd{HeLuo-2016} is still a classical Taylor expansion approach regardless of its analytical integrating skill (See \myRefInd{HeLuo-2016} for more detail).
The full equation in \myeqref{eq:CABKLBM} equals {\it exact} LB equation proposed by Boesh-Karlin approach\mycite{BoeschKarlin-2075}, denoted as Boesch-Karlin LB equation in this paper. Compared with Euler-Maclaurin integral in \myRefInd{BoeschKarlin-2075}, the derivation in ACI theory is mathematically more concise with a clear physical figure. }
In practice, the implicitness of $g^{n+1}$ would be removed by introducing an artificial distribution $h$, a trick proposed by \myRefInd{HeChen-173}.
Thus the implemented Boesch-Karlin LB equation is
\begin{equation}\label{eq:bkLB}
h^{n+1}-h^n=-\left(1-e^{-\Delta t/\lambda}\right)\left(h^n-g^n\right),
\end{equation}
where
\begin{equation}
\label{eq:bkExp}
h = f + \left( {\frac{{\Delta t}}{\lambda }\frac{1}{{1 - {e^{ - \Delta t/\lambda }}}} - 1} \right)\left( {f - g} \right).
\end{equation}
In computation, we deal $h$ as $f$, neglecting their difference. Then the practical Boesch-Karlin LB equation equals our aforementioned SA LB equation.\rvs{ As the relaxation time in \myeqref{eq:bkLB} is always $\le 1$, {\it Boesch and Karlin} assert that the overrelaxation in LBGK LB equation is beyond the kinetic theory of continuous-time BGK-Boltzmann equation\mycite{BoeschKarlin-2075}, which is questionable since the $g\left( {t'} \right)$ model behind Boesch-Karlin LB equation is quite rough. }

Now the question is what's the $g\left( {t'} \right)$ model behind LBGK equation. As we \mrvs{assume} that \mrvs{LB equation is an analytical characteristic integral} of BGK-Boltzmann equation and $g\left( {t'} \right)$  model ensures that the transient equation of particle interaction with respect to $\left( {\vec r,\vec \xi ,t} \right)$ is exact BGK-Boltzmann equation, then the $g\left(t'\right)$ model is not difficult to deduce, namely evolutionary correction of deviation (ECD) model, 
\begin{equation}\label{eq:ECDg}
g\left( {t'} \right) = \mrvs{f\left( {t'} \right)} - \frac{{t'}}{{\Delta t}}\left( {{f^{n + 1}} - {f^n}} \right) - {f^n} + {g^n}.
\end{equation}
\mrvs{It is emphasized that this equation is mathematical development for LBGK equation  under ACI theory.} Substituting it into the characteristic integral of BGK-Boltzmann equation, we reproduce LBGK equation,
\begin{equation}
{f^{n + 1}} - {f^n} =  - \frac{1}{{\lambda /\Delta t + 0.5}}\left( {{f^n} - {g^n}} \right).
\end{equation}
It worths noting that after substituting ECD model,  BGK-Boltzmann equation is integrable directly without requiring the introduce of integrating factor $e^{t/\lambda}$.
Similarly, the constant collision model which assumes that the collision changing rate along the characteristics is constant, namely direct correction of deviation (DCD) model, reads
\begin{equation}\label{eq:DCDg}
g\left( {t'} \right) = \mrvs{f\left( {t'} \right)} - {f^n} + {g^n},
\end{equation}
and its corresponding LB equation yields
\begin{equation}
{f^{n + 1}} - {f^n} =  - \frac{{\Delta t}}{\lambda }\left( {{f^n} - {g^n}} \right).
\end{equation}

We can also deduce the actual $g\left( {t'} \right)$  model behind the LB equation proposed by {\it He et al}\mycite{HeChen-173}, designated as He-Chen-Dong LB equation in this paper, \begin{equation}
g\left( {t'} \right) = \mrvs{f\left( {t'} \right)} - \frac{{t'}}{{\Delta t}}\left( {{f^{n + 1}} - {f^n}} \right) - {f^n} + {g^n}+ \frac{{t'}}{{\Delta t}}\left( {{g^{n + 1}} - {g^n}}\right),
\end{equation}
its LB equation reads
\begin{equation}
{f^{n + 1}} - {f^n} =  - \frac{{\Delta t}}{{2\lambda }}\left( {\left( {{f^n} - {g^n}} \right) + \left( {{f^{n + 1}} - {g^{n + 1}}} \right)} \right).
\end{equation}
Removing the implicitness of $g^{n + 1}$, the equation turns into
\begin{equation}
\label{eq:HCDLBM}
{h^{n + 1}} - {h^n} =  - \frac{1}{{0.5 + \lambda /\Delta t}}\left( {{h^n} - {g^n}} \right),
\end{equation} 
where
\begin{equation}
\label{eq:HCDExp}
h = f + \frac{{\Delta t}}{{2\lambda }}\left( {f - g} \right).
\end{equation}
Ignoring the difference between $h$ and $f$, He-Chen-Dong LB equation is equivalent with the standard LBGK equation (named ECD LB equation in this paper).

Summarizing all LB equations derived in this section, SA, Boesh-Karlin, ECD, DCD and He-Chen-Dong LB equations, they all are analytical characteristic integrals of BGK-Boltzmann equation. The difference \mrvs{is} their \mrvs{designs} of $g\left(t'\right)$ along the characteristics. In practice, Boesh-Karlin and He-Chen-Dong LB equation equal SA and ECD LB equation respectively.  All these LB equations are implemented with a unified form, 
\begin{equation}\label{eq:LB}
{f^{n + 1}} - {f^n} =  - \frac{1}{\tau }\left( {{f^n} - {g^n}} \right),
\end{equation}
with
\begin{equation}\label{eq:relax}
% MathType!Translator!2!1!LaTeX.tdl!LaTeX 2.09 and later!
\frac{1}{\tau } = \left\{ {\begin{array}{*{20}{l}}
	{1 - {e^{ - \Delta t/\lambda }},}&{{\rm{SA}}}\\
	{\frac{{\Delta t}}{\lambda },}&{{\rm{DCD}}}\\
	{1/\left( {0.5 + \frac{\lambda }{{\Delta t}}} \right),}&{{\rm{ECD}}}
	\end{array}} \right. .% MathType!End!2!1!
\end{equation}
It indicates that $1/\tau$, the relaxation time of LB method in classical LB theories\mycite{BoeschKarlin-2075,HeChen-173,Hwang-2023}, is actually a reflection of  $g\left(t'\right)$ model along the characteristics. 
It also worths noting that LB equation, the characteristic integral of BGK-Boltzmann equation, is by no means limited to the form in \myeqref{eq:LB}.

We close this section with a few remarks:
\begin{enumerate}[label={$\bullet$ }]  
	\item In this section, we propose ACI LB theory. In ACI LB theory, we take BGK-Boltzmann equation as the exact kinetic equation behind Navier-Stokes continuum and momentum equations and construct LB equation through rigorously integrating BGK-Boltzmann equation along characteristics. BGK-Boltzmann equation can be analytically integrated except Maxwell-Boltzmann distribution, which evolutes nonlinearly and needs to be calculated by LB equation. Thus the design of $g\left(t'\right)$, temporal evolution of equilibrium distribution along characteristics, decides the accuracy of LB equation. 
	\mrvs{In case the whole evolution of BGK-Boltzmann equation departs the physical process too much, we restrict the value of $g\left(t'\right)$ at $t'=0$ to $g^n$, ensuring that the transient equation of particle interaction with respect to $\left( {\vec r,\vec \xi ,t} \right)$ is exact BGK-Boltzmann equation.}

	\item In ACI LB theory, there are two important parameters, the collision time $\lambda$, depicting the time required for the distribution to reach equilibrium through collision, and the step time $\Delta t$, denoting the time interval of the LB equation. Their ratio, the collision number $\Delta t/ \lambda$, describes the times of the distribution reaching equilibrium state. It reflects the intensity of particle interaction in the time span of LB equation, independent from detailed $g\left(t'\right)$ models. By contrast, the traditionally assumed characteristic parameter of LB equation,  relaxation time, is merely a reflection of $g\left(t'\right)$ model. Thus the actual characteristic parameter of LB equation should be collision number $\Delta t/ \lambda$, which is supported by \myeqref{eq:relax}. 
	
	\item \rvs{ACI LB theory is a physically general theory. By tuning $g\left(t'\right)$ model, ACI LB theory can easily reproduce SA, Boesh-Karlin, ECD, DCD and He-Chen-Dong LB equations which are developed under different theories. 
	It makes the comparison of LB equations avoid the ambiguous theoretical analysis of their derivations, leads to a simple investigation of their designs of $g\left(t'\right)$ model. Another direct implication of this generality is that  the form of LB equation is by no means limited to \myeqref{eq:LB} since the flexibility of $g\left(t'\right)$ model is far more great than the presented ones in this section.}

	\item \rvs{ACI LB theory is a mathematically rigorous but concise theory. It avoids the viscosity inconsistence\mycite{HeChen-173} introduced by truncating Taylor series in Taylor expansion schemes\mycite{HeLuo-2016, SterlingChen-252}. 
	As an analytical characteristic integral approach like Boesch-Karlin approach, ACI theory eliminates the uncertainty of trapezoidal rule in He-Chen-Dong approach\mycite{HeChen-173} (See \myRefInd{BoeschKarlin-2075} for more detailed discussions). 
	Compared with Boesch-Karlin approach\mycite{BoeschKarlin-2075}, the integrating skill of ACI theory is mathematically more concise with a clear physical figure. And the asserted concept that $g\left(t'\right)$ model forges LB equation significantly extends the understanding of BGK-Boltzmann kinetic theory in Boesch-Karlin approach. 
	ACI theory also avoids the uncertainty of collision and streaming splitting operation in Strang splitting approach\mycite{Dellar-2074}. }

	\item \rvs{ACI LB theory is sensitive. The difference among LB equations will clearly manifest in their corresponding $g\left(t'\right)$ models. For example, the $g\left(t'\right)$ models corresponding to Boesh-Karlin and He-Chen-Dong LB equation are different from their implemented equations, i.e. SA and ECD LB equation. }

\end{enumerate}

\section{Discretization of LB equation }
\label{sec:disCrt}
In the former section, we have demonstrated the derivation of LB equation from BGK-Boltzmann equation. But the whole argument is based on velocity-continuous equation, which should be discretized before implementation.  In ACI LB theory, the discretization of LB equation in velocity space is independent from the construction of LB equation, it is directly performed on the BGK-Boltzmann equation. 
The discretion includes discretizing the velocity space of particle distribution and constructing the equilibrium distribution based on the discrete-velocity model.
Technically speaking, the discrete equilibrium distribution model determines the final recovery form of hydrodynamic equations, such as compressible or incompressible equations, conduct equation and so on, but it is out of this paper's discussion. Here we tend to offer a detail demonstration and illustrate the framework of ACI theory, i.e. the discretization of velocity space on BGK-Boltzmann equation, determining the practical LB algorithm, is independent from detailed LB equation. We take a classical discretization as the detail demonstration, small Mach-number approximation (SMA) approach, proposed by \myRefInd{HeLuo-2016}. For the sake of simplicity but without losing generality, we employ 2-dimensional Maxwell-Boltzmann distribution to construct the classical D2Q9 model. The derivation can be easily extended \mrvs{to 3 dimensions} and constructing new models.

The key idea of velocity discretization in ACI LB theory is to ensure that the assumed macroscopic hydrodynamic equations still can be recovered by the discretized BGK-Boltzmann equation. To achieve this, SMA approach employs keeping the relative velocity moment integrals of discretized equilibrium distribution consistent with Maxwell-Boltzmann distribution. For an instance, the following velocity moment integrals in \myeqref{eq:monInt} have been used in \mySecRef{sec:BGK} to recover the Navier-Stokes continuum and momentum equations, 
\begin{subequations}\label{eq:monInt}
\begin{align}
{\rho :}&~{1,{\xi _\alpha },{\xi _\alpha }{\xi _\beta }},\\
{\rho \vec u:}&~{{\xi _\alpha },{\xi _\alpha }{\xi _\beta },{\xi _\alpha }{\xi _\beta }{\xi _\gamma }},
\end{align}
\end{subequations}
where $\xi _\alpha$ is the component of $\vec \xi$ in Cartesian coordinates. So the discrete model should keep equivalent with Maxwell-Boltzmann distribution on moments $1,\vec \xi,\dots,\vec \xi ^3$.

To construct the discrete equilibrium distribution, SMA approach firstly decomposes the exponent part in the Maxwell-Boltzmann distribution: one for the weight function, the other for the distribution with respect to macro velocity,
\begin{equation}\label{eq:dcomMB}
% MathType!Translator!2!1!LaTeX.tdl!LaTeX 2.09 and later!
g = \frac{\rho }{{{{\left( {2\pi RT} \right)}^{D/2}}}}\exp \left( { - \frac{{{\xi ^2}}}{{2RT}}} \right)\exp \left( { - \frac{{{u^2} - 2\vec u \cdot \vec \xi }}{{2RT}}} \right).% MathType!End!2!1!
\end{equation}

For the convenience of calculating weights, the truncated small velocity expansion (or small-Mach-number approximation) is employed\mycite{HeLuo-2016}, where the terms above 2nd macro-velocity order are neglected,
\begin{align}\label{eq:MBAprx}
% MathType!Translator!2!1!LaTeX.tdl!LaTeX 2.09 and later!
g' = &\frac{\rho }{{{{\left( {2\pi RT} \right)}^{D/2}}}}\exp \left( { - \frac{{{\xi ^2}}}{{2RT}}} \right) \nonumber \\
&\times  \left( {1 + \frac{{\vec \xi  \cdot \vec u}}{{RT}} + \frac{{{{\left( {\vec \xi  \cdot \vec u} \right)}^2}}}{{2{{\left( {RT} \right)}^2}}} - \frac{{{u^2}}}{{2RT}}} \right) + O\left( {{u^3}} \right).% MathType!End!2!1!
\end{align}
It worths noting that the order of small-Mach-number approximation decides the velocity moment integrals' accuracy of $g'$, i.e. the highest order of velocity moment integral that $g'$ can restore. Taking \myeqref{eq:MBAprx} for example, the 2nd order truncation could only restore the velocity moment integral up to 2nd order. The 3rd order moment integral of the truncated Maxwell-Boltzmann distribution in \myeqref{eq:dMBMom3d} would deviate from the original in \myeqref{eq:MBMom3d},
\begin{align}\label{eq:MBMom3d}
% MathType!Translator!2!1!LaTeX.tdl!LaTeX 2.09 and later!
\int {g{\xi _\alpha }{\xi _\beta }{\xi _\gamma }}  = &\rho {u_\alpha }{u_\beta }{u_\gamma } + {u_\gamma }\rho RT{\delta _{\alpha \beta }} \nonumber \\
&+ {u_\beta }\rho RT{\delta _{\alpha \gamma }} + {u_\alpha }\rho RT{\delta _{\beta \gamma }},% MathType!End!2!1!
\end{align}
\begin{equation}\label{eq:dMBMom3d}
% MathType!Translator!2!1!LaTeX.tdl!LaTeX 2.09 and later!
\int {g'{\xi _\alpha }{\xi _\beta }{\xi _\gamma }}  = {u_\gamma }\rho RT{\delta _{\alpha \beta }} + {u_\beta }\rho RT{\delta _{\alpha \gamma }} + {u_\alpha }\rho RT{\delta _{\beta \gamma }}.% MathType!End!2!1!
\end{equation}
To keep equivalent on 3rd velocity moment integral, $g'$ needs to maintain macro-velocity terms up to 3rd order. Here are 3rd order approximation (seeing \myeqref{eq:MBAprx3}) and its 3rd velocity moment integral (seeing \myeqref{eq:dMBMom3dInt}),
\begin{equation}\label{eq:MBAprx3}
% MathType!Translator!2!1!LaTeX.tdl!LaTeX 2.09 and later!
g' = \frac{\rho }{{{{\left( {2\pi RT} \right)}^{D/2}}}}\exp \left( { - \frac{{{\xi ^2}}}{{2RT}}} \right) \left( {1 + \frac{{2\vec \xi \cdot \vec u - {u^2}}}{{2RT}} + \frac{1}{2}{{\left( {\frac{{\vec \xi \cdot \vec u}}{{RT}}} \right)}^2} + \frac{1}{2}\left( {\frac{{ - \vec \xi \cdot {{\vec u}^3}}}{{{{\left( {RT} \right)}^2}}}} \right) + \frac{1}{6}{{\left( {\frac{{\vec \xi \cdot \vec u}}{{RT}}} \right)}^3}} \right),% MathType!End!2!1!
\end{equation}
\begin{align}\label{eq:dMBMom3dInt}
% MathType!Translator!2!1!LaTeX.tdl!LaTeX 2.09 and later!
\int {g'{\xi _\alpha }{\xi _\beta }{\xi _\gamma }}  = &\rho {u_\alpha }{u_\beta }{u_\gamma } + {u_\gamma }\rho RT{\delta _{\alpha \beta }} \nonumber \\
&+ {u_\beta }\rho RT{\delta _{\alpha \gamma }} + {u_\alpha }\rho RT{\delta _{\beta \gamma }}.% MathType!End!2!1!
\end{align}
Unfortunately, the implementation of 3rd order approximation requires \mrvs{a more complicated discrete velocity set}. In this paper, we still employ the 2nd-order approximation.

After small-Mach-number approximation, the target continuous equilibrium distribution has been $g'$ instead of original Maxwell-Boltzmann distribution $g$. Right now, $g'$ in \myeqref{eq:MBAprx} bears a strong resemblance to the classical D2Q9 equilibrium distribution, all remains to be accomplished is calculating the weight of discrete velocity. As we discussed before, the discrete equilibrium model $f^{eq}$ should keep the velocity moment integrals required in \myeqref{eq:monInt} consistent with $g'$,
\begin{equation}\label{eq:momCons}
% MathType!Translator!2!1!LaTeX.tdl!LaTeX 2.09 and later!
\sum {\psi \left( {{{\vec \xi }_\alpha }} \right)f_\alpha ^{eq}}  = \int {\psi \left( {\vec \xi } \right)g'd\vec \xi }, % MathType!End!2!1!
\end{equation}
where $\psi \left( {\vec \xi } \right)$ is a polynomial of $\vec \xi $ up to order 3. For the sake of simplicity but without losing generality, assuming
\begin{equation}\label{eq:momExp}
% MathType!Translator!2!1!LaTeX.tdl!LaTeX 2.09 and later!
{\psi \left( {\vec \xi } \right) = \psi _{m,n}}\left( {\vec \xi } \right) = \xi _x^m\xi _y^n,m + n \le 3,% MathType!End!2!1!
\end{equation}
then the velocity moment integrals of $g'$ can read as
\begin{equation}\label{eq:appMomInt}
% MathType!Translator!2!1!LaTeX.tdl!LaTeX 2.09 and later!
\begin{array}{l}
\int {\psi \left( {\vec \xi } \right)g'd\vec \xi }
= \frac{\rho }{\pi }\left( \begin{array}{l}
\left( {1 - \frac{{{u^2}}}{{2RT}}} \right){\left( {\sqrt {2RT} } \right)^{m + n}}{I_m}{I_n} + \frac{1}{{RT}}{\left( {\sqrt {2RT} } \right)^{m + n + 1}}\left( {{u_x}{I_{m + 1}}{I_n} + {u_y}{I_m}{I_{n + 1}}} \right)\\
+ \frac{{\text{1}}}{{2{{\left( {RT} \right)}^2}}}{\left( {\sqrt {2RT} } \right)^{m + n + 2}}\left( {{I_{m + 2}}{I_n}u_x^2 + 2{I_{m + 1}}{I_{n + 1}}{u_x}{u_y} + {I_m}{I_{n + 2}}u_y^2} \right)
\end{array} \right)
\end{array},% MathType!End!2!1!,
\end{equation}
where
\begin{equation}\label{eq:expInt}
% MathType!Translator!2!1!LaTeX.tdl!LaTeX 2.09 and later!
{I_m} = \int {{e^{ - {x^2}}}{x^m}dx} ,x = \xi /\sqrt {2RT},% MathType!End!2!1!
\end{equation}
which can be calculated numerically with Gaussian-type quadrature. Our object is using proper discretization of velocity space to evaluate the integral $I_m$ with $m$ up to 5. Naturally, the third-order Hermite formula\mycite{HeLuo-2016} is the optimal choice for the purpose of deriving the D2Q9 model,
\begin{equation}\label{eq:HermF}
% MathType!Translator!2!1!LaTeX.tdl!LaTeX 2.09 and later!
{I_m} = \int {{{e }^{ - {x^2}}}{x^m}dx}  = \sum {{\omega _i}x_i^m}. % MathType!End!2!1!
\end{equation}
The integrals need to be evaluated are, $I_0$, $I_2$, $I_4$. Other integrals with odd \mrvs{orders} of $x$ equal 0 due to the symmetry of \myeqref{eq:HermF} which can be easily solved by using symmetrical abscissas of the quadrature. Setting the abscissas of quadrature as $x_0=0$,$x_1=-\zeta $, $x_2=\zeta $, we have the following three equations:
\begin{subequations}\label{eq:evInt}
	\begin{equation}
	{I_0} = {\omega _0} + {\omega _1} + {\omega _2} = \sqrt \pi,
	\end{equation}
	\begin{equation}
	{I_2} = {\omega _1}{\zeta ^2} + {\omega _2}{\zeta ^2} = \sqrt \pi  /2,
	\end{equation}
	\begin{equation}
	{I_4} = {\omega _1}{\zeta ^4} + {\omega _2}{\zeta ^4} = 3\sqrt \pi  /4,
	\end{equation}
\end{subequations}
with symmetrical weights
\begin{equation}\label{eq:symWght}
\omega _1 = \omega _2.
\end{equation}
The three abscissas of the quadrature are
\begin{equation}\label{eq:quaAbs}
\begin{array}{*{20}{c}}
{{x_0} = 0,}&{{x_1} =  - \sqrt {3/2} ,}&{{x_1} = \sqrt {3/2} },
\end{array}
\end{equation}
and the corresponding weight coefficients are
\begin{equation}\label{eq:wegt}
\begin{array}{*{20}{c}}
{{\omega _0} = 2\sqrt \pi  /3,}&{{\omega _1} = \sqrt \pi  /6,}&{{\omega _1} = \sqrt \pi  /6},
\end{array}
\end{equation}
Then the evaluation \myeqref{eq:momCons} can read as
\begin{equation}\label{eq:DmomCons}
\sum {\psi \left( {{{\vec \xi }_\alpha}} \right)f_\alpha ^{eq}}  = \int {\psi \left( {\vec \xi } \right)g'd\vec \xi }
= \frac{\rho }{\pi }\sum\limits_{i ,j  = 0}^2 {{\omega _i }{\omega _j }\psi \left( {{\vec \xi _{i ,j }}} \right)\left( {1 + \frac{{{{\vec \xi }_{i ,j }} \cdot \vec u}}{{RT}} + \frac{{{{\left( {{{\vec \xi }_{i ,j }} \cdot \vec u} \right)}^2}}}{{2{{\left( {RT} \right)}^2}}} - \frac{{{u^2}}}{{2RT}}} \right)},
\end{equation}
where ${{\vec \xi }_\alpha} = {{\vec \xi }_{i ,j }} = \sqrt {2RT} \left( {{x_i },{x_j }} \right)$.  It's straightforward to identify the discrete equilibrium distribution with
\begin{equation}\label{eq:disEq}
f_\alpha^{eq} = \frac{\rho}{\pi }{\omega _i }{\omega _j }\left( {1 + \frac{{{{\vec \xi }_{i ,j }} \cdot \vec u}}{{RT}} + \frac{{{{\left( {{{\vec \xi }_{i ,j }} \cdot \vec u} \right)}^2}}}{{2{{\left( {RT} \right)}^2}}} - \frac{{{u^2}}}{{2RT}}} \right).
\end{equation}
Combining with the notations of discrete velocity and corresponding weights, employing the relation $RT = c_s^2 = c^2/3$, a complete D2Q9 model is constructed,
\begin{equation}\label{eq:D2Q9}
f_\alpha ^{eq} = {w_\alpha }\rho \left( {1 + \frac{{{{\vec \xi }_\alpha } \cdot \vec u}}{{c_s^2}} + \frac{{{{\left( {{{\vec \xi }_\alpha } \cdot \vec u} \right)}^2}}}{{2c_s^4}} - \frac{{{u^2}}}{{2c_s^2}}} \right),
\end{equation}
with discrete velocity,
\begin{equation}\label{eq:dVel}
{\vec \xi _\alpha } = \left\{ {\begin{array}{*{20}{l}}
	{\left( {0,0} \right),}&{\alpha  = 0}\\
	{\left( {\cos {\theta _\alpha },\sin {\theta _\alpha }} \right)c,}&{\alpha  = 1,2,3,4}\\
	{\sqrt 2 \left( {\cos {\theta _\alpha },\sin {\theta _\alpha }} \right)c,}&{\alpha  = 5,6,7,8}
	\end{array}} \right.,
\end{equation}
where
\[ {\theta _\alpha } = \left\{ {\begin{array}{*{20}{l}}
	{\left( {\alpha  - 1} \right)\pi /2,}&{\alpha  = 1,2,3,4}\\
	{\left( {\alpha  - 5} \right)\pi /2 + \pi /4,}&{\alpha  = 5,6,7,8}
	\end{array}} \right., \]
and weights,
\begin{equation}\label{eq:Dwght}
{w_\alpha } = \frac{{{\omega _i}{\omega _j}}}{\pi } = \left\{ {\begin{array}{*{20}{l}}
	{4/9,}&{\alpha  = 0}\\
	{1/9,}&{\alpha  = 1,2,3,4}\\
	{1/36,}&{\alpha  = 5,6,7,8}
	\end{array}} \right.,
\end{equation}
where $\rho  = \sum {{f_\alpha }}$, $\vec u = \sum {{f_\alpha }{{\vec \xi }_\alpha }} /\rho $.

The derived D2Q9 model is directly constructed on BGK-Boltzmann equation, independent from detail LB equation. Thus it is applicable \mrvs{to} all $g\left(t'\right)$ models.

\section{Numerical analysis}
\label{sec:numRe}
Within ACI LB theory, all popular LB equations are analytical characteristic integrals of BGK-Boltzmann equation, identified by their designs of $g\left(t'\right)$ model, then all we need  to analyze is their designs of $g\left(t'\right)$ model.  
ACI LB theory also proposes a new characteristic parameter of LB equation, collision number $\Delta t/\lambda$, depicting the particle-interacting intensity in the time span of LB equation.
Unlike the traditionally assumed characteristic parameter, relaxation time $1/\tau$, which is merely a reflection of $g\left(t'\right)$ model,  the collision number is independent from detail $g\left(t'\right)$ models. In this section, we would analyze the numerical performance of derived LB equations under different collision numbers. 
In numerical computation,  \mrvs{Boesh-Karlin and He-Chen-Dong LB equation would be replaced by SA and ECD LB equation, ignoring the implicitness as their proposers suggested\mycite{BoeschKarlin-2075,HeChen-173}.} We start with the comparison of  relaxation time curves along collision number as \mrvs{all derived LB equations} share a unified equation form in \myeqref{eq:LB}, only distinguished in relaxation time formulas. Then two benchmarks, unsteady Couette flow and lid-driven cavity flow, will be calculated by SA and ECD LB equations, employing the derived D2Q9 model. The results of SA and ECD will be compared to analyze their accuracy under different collision numbers and validate the effect of $g\left(t'\right)$ model. In the calculation of benchmarks, the DCD approximation will be abandoned due to its instability.

\subsection{\mrvs{Relaxation} time}
\label{subsec:retim}
\begin{figure}[htbp]
	\centering
	\includegraphics[width=0.4\textwidth]{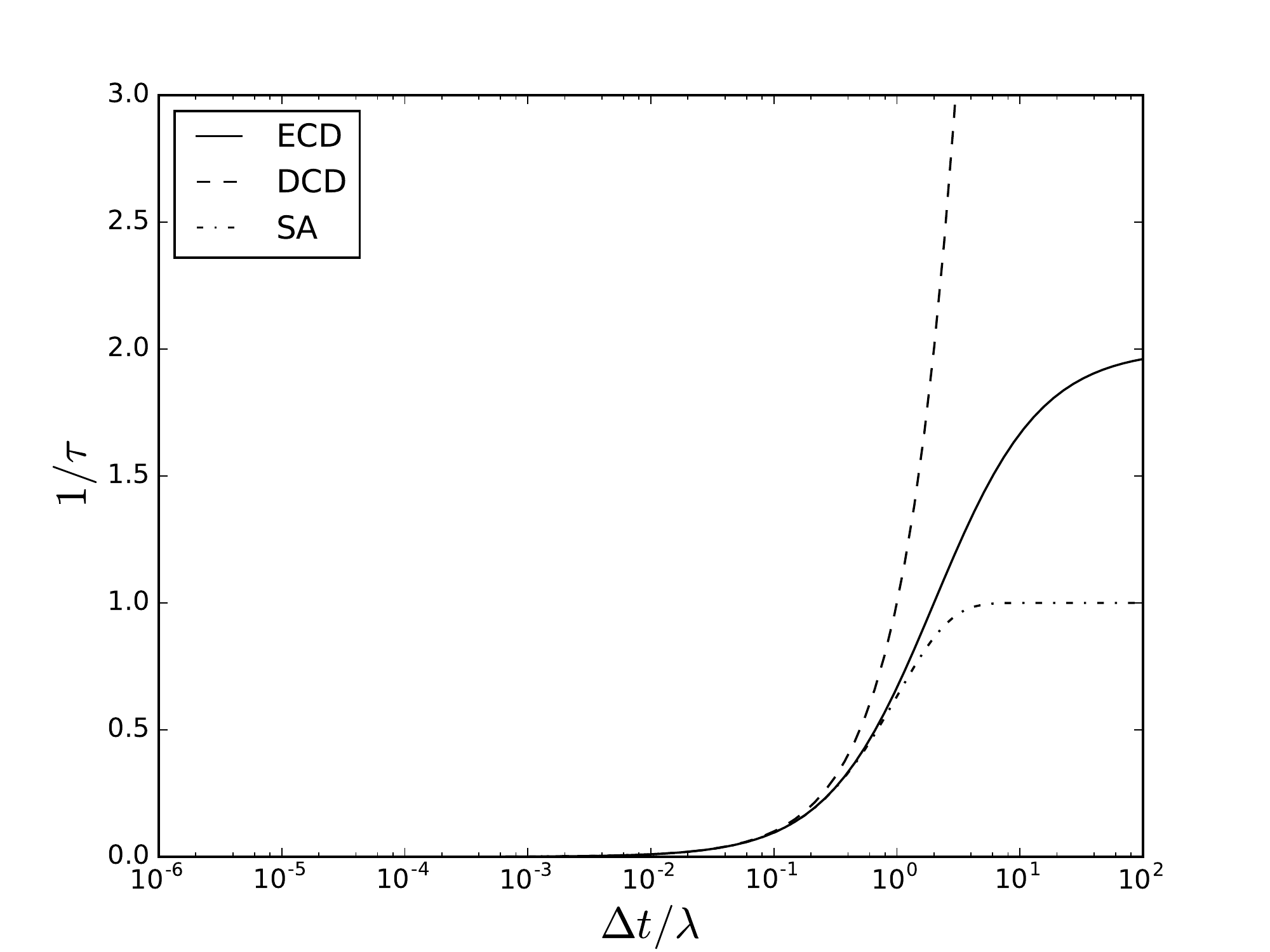}
	\caption{relaxation time vs collision number. the solid, dashed and dashdot \mrvs{lines are} ECD, DCD and SA result, respectively. Note that the result of DCD is truncated by the limitation of relaxation time. }\label{fig:rexVsKn}
\end{figure}

As \myeqref{eq:relax} shows, for the same calculation, i.e. the same collision number $\Delta t/\lambda $, the proposed $g\left(t'\right)$  models, SA, DCD and ECD,  only differ in relaxation time $1/\tau $. Fig.~\ref{fig:rexVsKn} illustrates the relaxation time profiles of SA, DCD and ECD with respect to collision number. For $\Delta t/\lambda \to 0$, SA, DCD and ECD approaches each other. As the collision number increases, all three relaxation times increase with different rate, which leads to $1/{\tau _{DCD}} > 1/{\tau _{ECD}} > 1/{\tau _{SA}}$. When the collision number goes into infinite, SA and ECD relaxation time would converge to 1 and 2 respectively, while DCD increase unlimitedly, which causes the instability of DCD.  Fig.~\ref{fig:rexVsKn} indicates that SA, DCD and ECD LB equations would have same numerical performance under low collision number meanwhile distinguish from each other under high collision number. This conclusion is supported by the design of  $g\left(t'\right)$. Once collision number is small enough, which \mrvs{indicates} small collision change, the particle distribution \mrvs{$f\left(t'\right)$} along characteristics would not depart $f^n$ too much. Then the modification parts in \mrvs{ECD model \myeqref{eq:ECDg} and DCD model \myeqref{eq:DCDg} would both approach 0},
\begin{subequations}\label{eq:modLCN}
	\begin{align}
	{\mrvs{f\left(t'\right)} - \frac{{t'}}{{\Delta t}}\left( {{f^{n + 1}} - {f^n}} \right) - {f^n} \cong 0},&\qquad {\text{ECD}}, \\
	{\mrvs{f\left(t'\right)}-f^n \cong 0},&\qquad {\text{DCD}}.
	\end{align}
\end{subequations}
Hence the $g\left(t'\right)$ models of ECD and DCD can be approximated as
\begin{equation}
g\left(t'\right)=g^n,
\end{equation}
which is exactly SA model. When the collision change increases as the collision number increases, the modification \mrvs{parts} in \myeqref{eq:modLCN} \mrvs{are} not neglectable. Thus SA, ECD and DCD model would show different results.  It worths noting that under large collision change, SA and DCD model are theoretically unsuitable as SA takes the equilibrium distribution constant and DCD takes the collision change rate constant, which contradict the physical process. What's more, Fig.~\ref{fig:rexVsKn} also indicates the over relaxation time problem is merely a manifestation of $g\left(t'\right)$  model under high collision number, irrelevant to particle kinetics.  

\subsection{Unsteady Couette flow}
\label{subsec:couet}
Unsteady Couette flow, also as known as transient plane Couette flow, is a typical numerical benchmark. It evolves in a straight channel, which infinitely extends in the x direction. The walls of the channel are parallel to the x axis and defined by the equations $y=0$ for the lower wall and $y=L$  for the upper wall. The flow is stationary at the beginning and driven by the upper wall with a constant velocity  after that. Its equation can be described as
\begin{equation}\label{eq:coetCtrl}
% MathType!Translator!2!1!LaTeX.tdl!LaTeX 2.09 and later!
\frac{{\partial u\left( {y,t} \right)}}{{\partial t}} = \nu \frac{{{\partial ^2}u\left( {y,t} \right)}}{{\partial {y^2}}},% MathType!End!2!1!
\end{equation}
with  the following initial and boundary conditions in the range of % MathType!Translator!2!1!LaTeX.tdl!LaTeX 2.09 and later!
$0 \le y \le L$,% MathType!End!2!1!
\begin{equation}\label{eq:coetCond}
% MathType!Translator!2!1!LaTeX.tdl!LaTeX 2.09 and later!
\begin{array}{*{20}{c}}
{u\left( {0,t} \right) = 0,}&{u\left( {L,t} \right) = U,}&{u\left( {0,t} \right) = 0},
\end{array}% MathType!End!2!1!
\end{equation}
where % MathType!Translator!2!1!LaTeX.tdl!LaTeX 2.09 and later!
$\nu $ % MathType!End!2!1!
is kinematic viscosity, $U$ is the upper wall's driven velocity, $L$ is the width of channel. Unsteady Couette flow can be solved analytically:
\begin{equation}\label{eq:coueSlo}
 % MathType!Translator!2!1!LaTeX.tdl!LaTeX 2.09 and later!
u\left( {y,t} \right) = U\frac{y}{L} - \frac{{2U}}{\pi }\sum\limits_{k = 1}^\infty  {\frac{1}{k}\sin \left( {k\pi \left( {1 - \frac{y}{L}} \right)} \right)} {e^{ - \frac{{{k^2}{\pi ^2}}}{{{L^2}}}\nu t}}.% MathType!End!2!1!
\end{equation}

\begin{figure}[htbp]
	\centering
	\begin{subfigure}[htbp]{0.4\textwidth}
		\caption{\label{fig:lcn}Low collision number}
		\includegraphics[width=\textwidth]{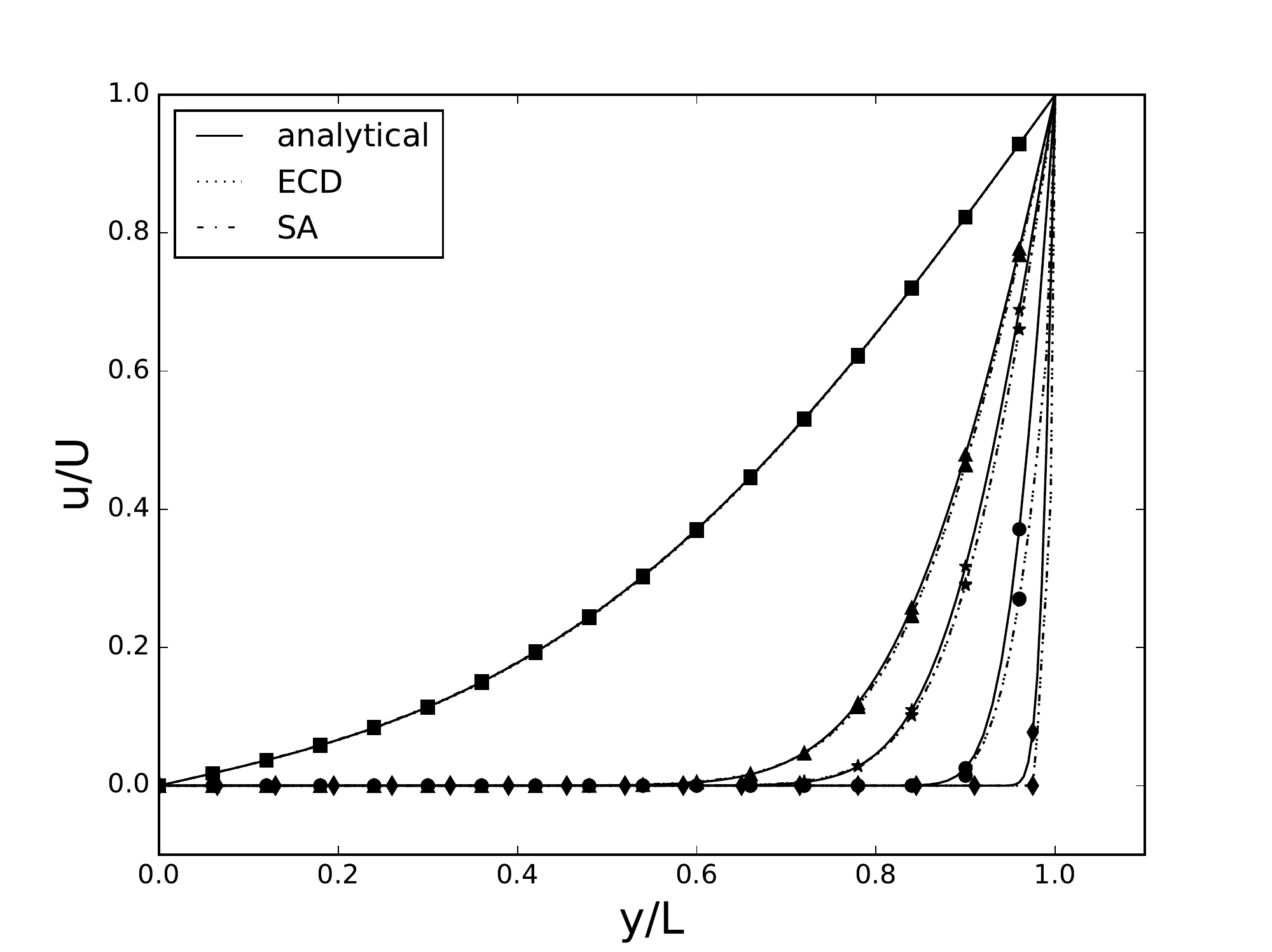}
	\end{subfigure}
	\begin{subfigure}[htbp]{0.4\textwidth}
		\caption{\label{fig:hcn}High collision number}
		\includegraphics[width=\textwidth]{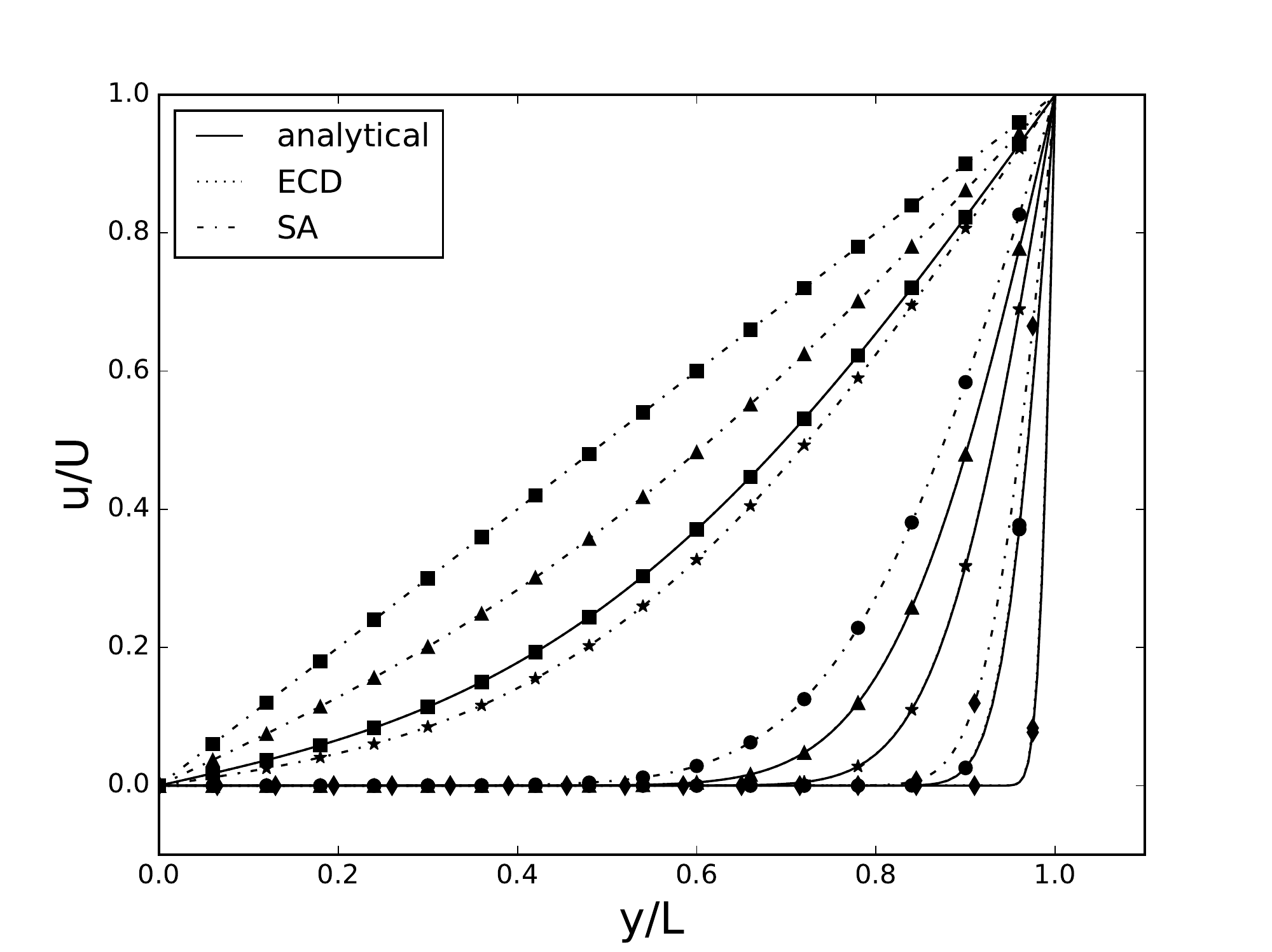}
	\end{subfigure}
	\caption{\label{fig:Couet}The velocity profiles of unsteady Couette flow at $t=1.00E-3\; \rm{s}$: (a) low collision number; (b) high collision number. The symbols $\blacksquare$, $\blacktriangle$, $\star$, $\bullet$, $\blacklozenge$ are results of case1,case2,case3,case4,case5, respectively. Under low collision number in Fig.~\ref{fig:lcn}, models agree well with each other, and share a similar profile with analytical solution. Under high collision number in Fig.~\ref{fig:hcn}, ECD keeps consistent with analytical solution while SA fails.
		Note the detail descriptions of cases are listed in Table~\ref{tab:unstdCase}.}
\end{figure}

\begin{figure}[htbp]
	\centering
	\begin{subfigure}[htbp]{0.4\textwidth}
		\caption{Velocity error evolution of case1}
		\includegraphics[width=\textwidth]{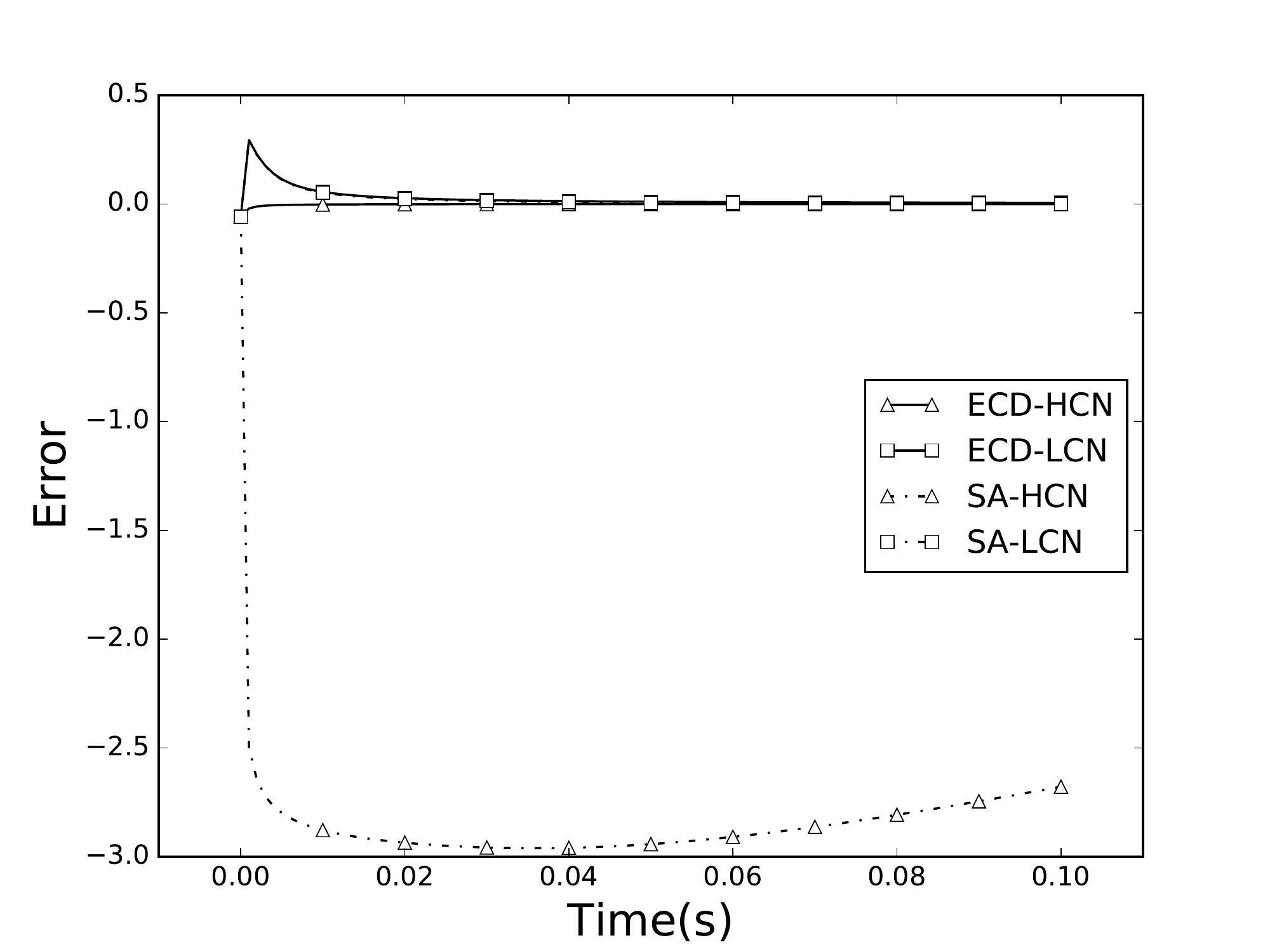}
	\end{subfigure}
	\begin{subfigure}[htbp]{0.4\textwidth}
		\caption{Velocity error evolution of case3}
		\includegraphics[width=\textwidth]{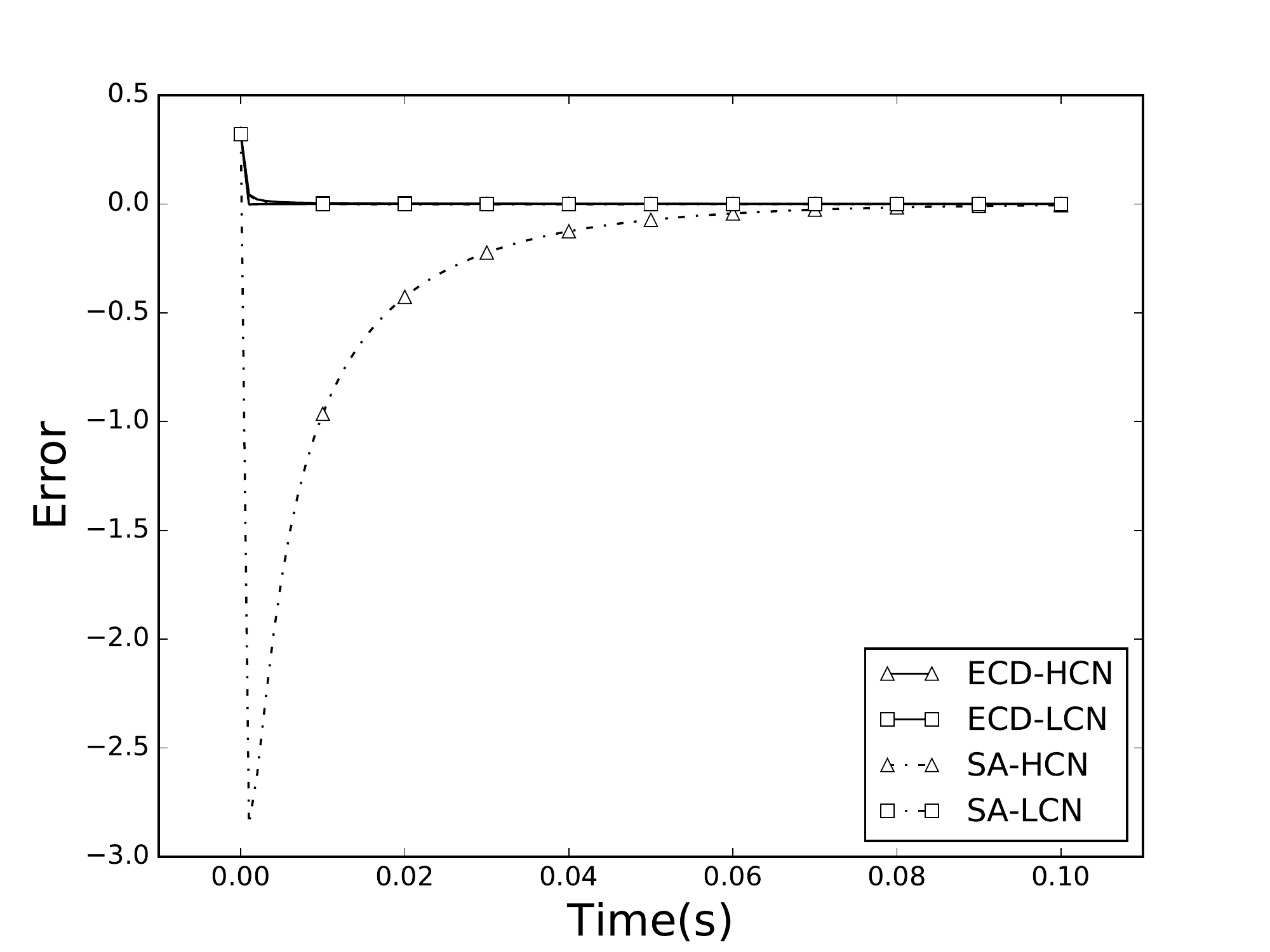}
	\end{subfigure}
	\begin{subfigure}[htbp]{0.4\textwidth}
		\caption{Velocity error evolution of case5}
		\includegraphics[width=\textwidth]{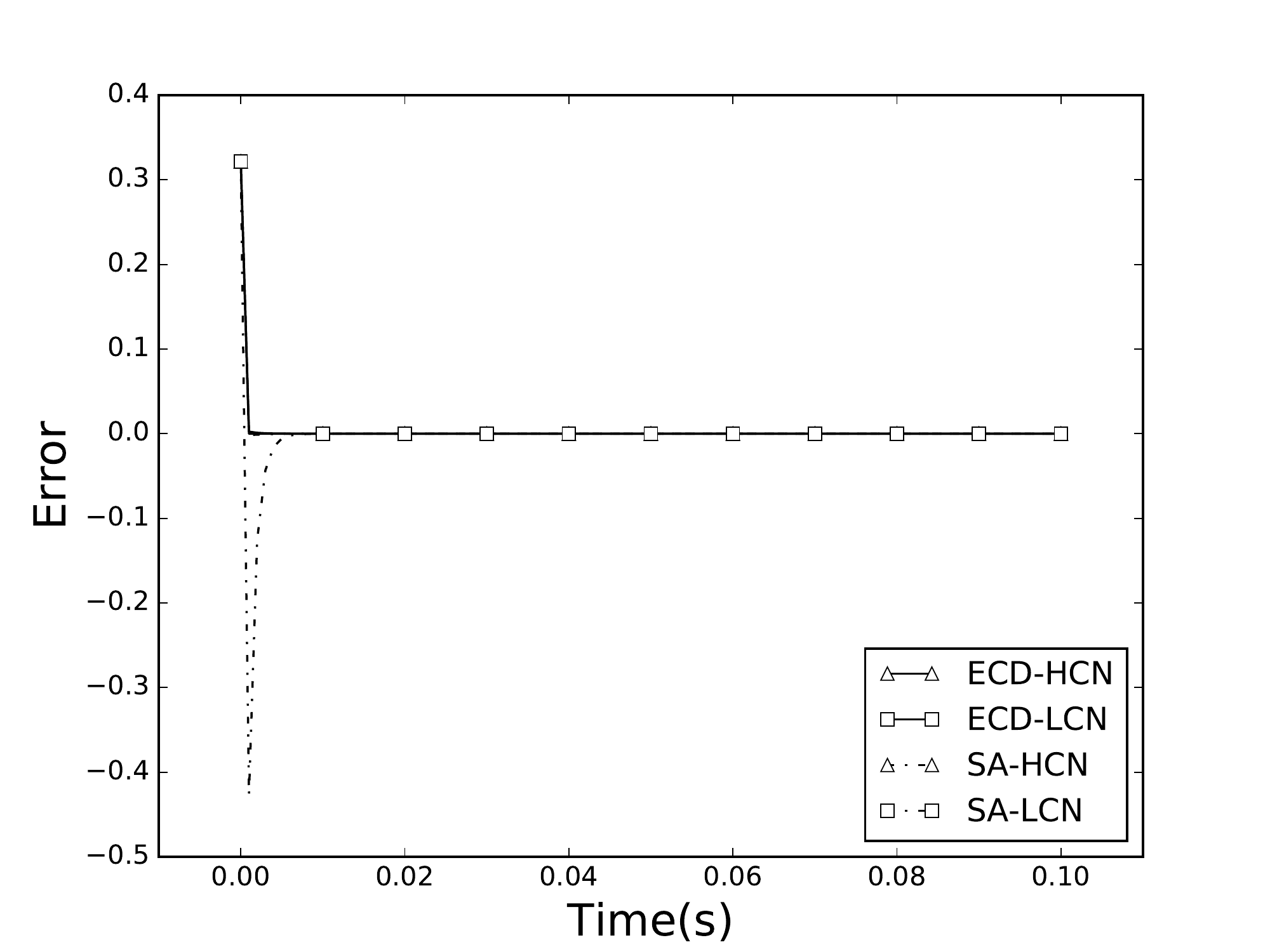}
	\end{subfigure}
	\caption{Velocity error evolution of unsteady Couette flow: (a) case1 result; (b) case3 result; (c) case5 result. ECD-HCN, ECD-LCN, SA-HCN, SA-LCN stand for results of ECD model with low collision number(LCN), ECD model with high collision number(HCN), SA model with low collision number(LCN), SA model with high collision number(HCN), respectively. }\label{fig:CouetErr}
\end{figure}

The periodic boundary is implemented on x direction to simulate the infinite extension. For the upper and lower wall, the regularized boundary \myRefInd{LattChopard-124} is applied. Actually, in unsteady Couette flow, Zou-He boundary\mycite{ZouHe-251} and Inamuro boundary\mycite{InamuroYoshino-238} would share the same macro result with regulariezed boundary, though the particle distributions $f$ may differ. 5 cases are designed to discuss the SA and ECD models \mrvs{(See Table~\ref{tab:unstdCase})}. Each case has two states: low collision number(LCN) and high collision number(HCN). Fig.~\ref{fig:Couet} shows the SA and ECD velocity profiles of unsteady Couette flow at $t=1.00E-3\; \rm{s}$. Fig.~\ref{fig:CouetErr}
 illustrates the error evolution on time. The error formula is defined as
\begin{equation}\label{eq:couetErr}
% MathType!Translator!2!1!LaTeX.tdl!LaTeX 2.09 and later!
Error = \frac{{\sum {\left( {u - {u_{ana}}} \right)} }}{{\sum {{u_{ana}}} }},% MathType!End!2!1!
\end{equation}
where $u$ is the computing velocity profile,  $u_{ana}$ is the analytical velocity profile. \mrvs{For the sake of space saving, only the error evolutions of case1,case3 and case5 are plotted. }

\begin{table}[htbp]
	\centering
	\caption{Configurations of unsteady Couette flow demonstrating cases. For all cases, the channel width equals $L = 0.1\; \rm{m}$, and the driven velocity of upper wall is $U = 1.0 \; \rm{m/s}$.}\label{tab:unstdCase}
	\begin{tabular}{l ccccc}
		\hline\hline
		\noalign{\smallskip}
		Case & CN\footnote{CN stands for collision number} & Grid &  & Kinematic & Collision \\[-1ex]
		name & type & number\footnote{Grid number is the mesh number on cavity side length $L$} & \raisebox{1.5ex}{Timestep} & viscosity & number \\
		\noalign{\smallskip}
		\hline
		% after \\: \hline or \cline{col1$-$col2} \cline{col3$-$col4} ...
		\noalign{\smallskip}
		& LCN &  & 2.50E$-$4 &  & 3.33E$-$1 \\[-1ex]
		\raisebox{1.5ex}{case1}& HCN & \raisebox{1.5ex}{201} & 2.50E$-$6 & \raisebox{1.5ex}{1.00E$-$3} & 3.33E$+$1 \\[0.5ex]
		& LCN &  & 1.00E$-$4 &  & 3.33E$-$1 \\[-1ex]
		\raisebox{1.5ex}{case2}& HCN & \raisebox{1.5ex}{101} & 1.00E$-$6 & \raisebox{1.5ex}{1.00E$-$2} & 3.33E$+$1 \\[0.5ex]
		& LCN &  & 2.00E$-$5 &  & 3.33E$-$1 \\[-1ex]
		\raisebox{1.5ex}{case3}& HCN & \raisebox{1.5ex}{101} & 2.00E$-$7 & \raisebox{1.5ex}{5.00E$-$2} & 3.33E$+$1 \\[0.5ex]
		& LCN &  & 1.00E$-$5 &  & 3.33E$-$1 \\[-1ex]
		\raisebox{1.5ex}{case4}& HCN & \raisebox{1.5ex}{101} & 1.00E$-$7 & \raisebox{1.5ex}{1.00E$-$1} & 3.33E$+$1 \\[0.5ex]
		& LCN &  & 1.00E$-$6 &  & 3.33E$-$1 \\[-1ex]
		\raisebox{1.5ex}{case5}& HCN & \raisebox{1.5ex}{101} & 1.00E$-$8 & \raisebox{1.5ex}{1.00E$+$0} & 3.33E$+$1 \\
		\noalign{\smallskip}
		\hline\hline
	\end{tabular}
\end{table}


\rvs{The velocity profiles of SA LB equation in Fig.~\ref{fig:Couet} show that at LCN, SA results are close to the analytical solutions meanwhile at HCN, they evidently deviate, manifesting magnified viscosities. By contrast, ECD results always keep consistent with analytical solutions regardless of collision number. The deviation of SA LB equation under HCN is explainable, due to the intensive collision change, the equilibrium distribution would be greatly changed, which leads to the failure of the constant equilibrium distribution assumption in SA model. And the results of ECD LB equation indicate that the modification part in ECD model significantly improves its numerical performance under HCN. Comparing the results of SA and ECD LB equation, it clearly shows that $g\left(t'\right)$ model determines the accuracy of LB equation.
The consistent numerical performance of SA and ECD LB equation under different collision numbers also validates that the collision number is a good characteristic parameter to depict LB equation. It reflects the collision intensity in the time span of LB equation, independent from detail $g\left(t'\right)$ model. }

\subsection{Lid-driven cavity flow}
\label{subsec:lidCav}
Lid-driven cavity flow is a classical constant-property benchmark for fluid computation, for its simple geometry and complicated flow behaviors, especially the corner flow phenomena. Lid-driven cavity flow is a 2-D case characterized by Renold (Re) number, which develops in a square cavity with the side length $L$. The fluid would keep stationary at the beginning and be driven by the top lid with constant velocity  after that. Except the top lid, all other walls keep rest. This case has no analytic solution.
As a convention, the results in \myRefInd{GhiaGhia-119} are taken as a calibration on velocity profiles.
As the definition of collision number shows, with small kinematic viscosity, the collision number is inherently large.
It's very difficult to simulate a high Re-number cavity flow under low collision number without mesh explosion. Fortunately, our mission is to evaluate $g\left(t'\right)$ models under collision number instead of Re number. A small Re-number case with different collision numbers will fulfill the task.
Hence the case with smallest Re number in \myRefInd{GhiaGhia-119} , $\rm{Re}=100$, is calculated for analysis. 

\begin{figure}[htbp]
	\centering
	\begin{subfigure}[htbp]{0.4\textwidth}
		\caption{\label{fig:CavVln}Velocity profiles of $U_x$ along the vertical line through cavity center}%
		\includegraphics[width=\textwidth]{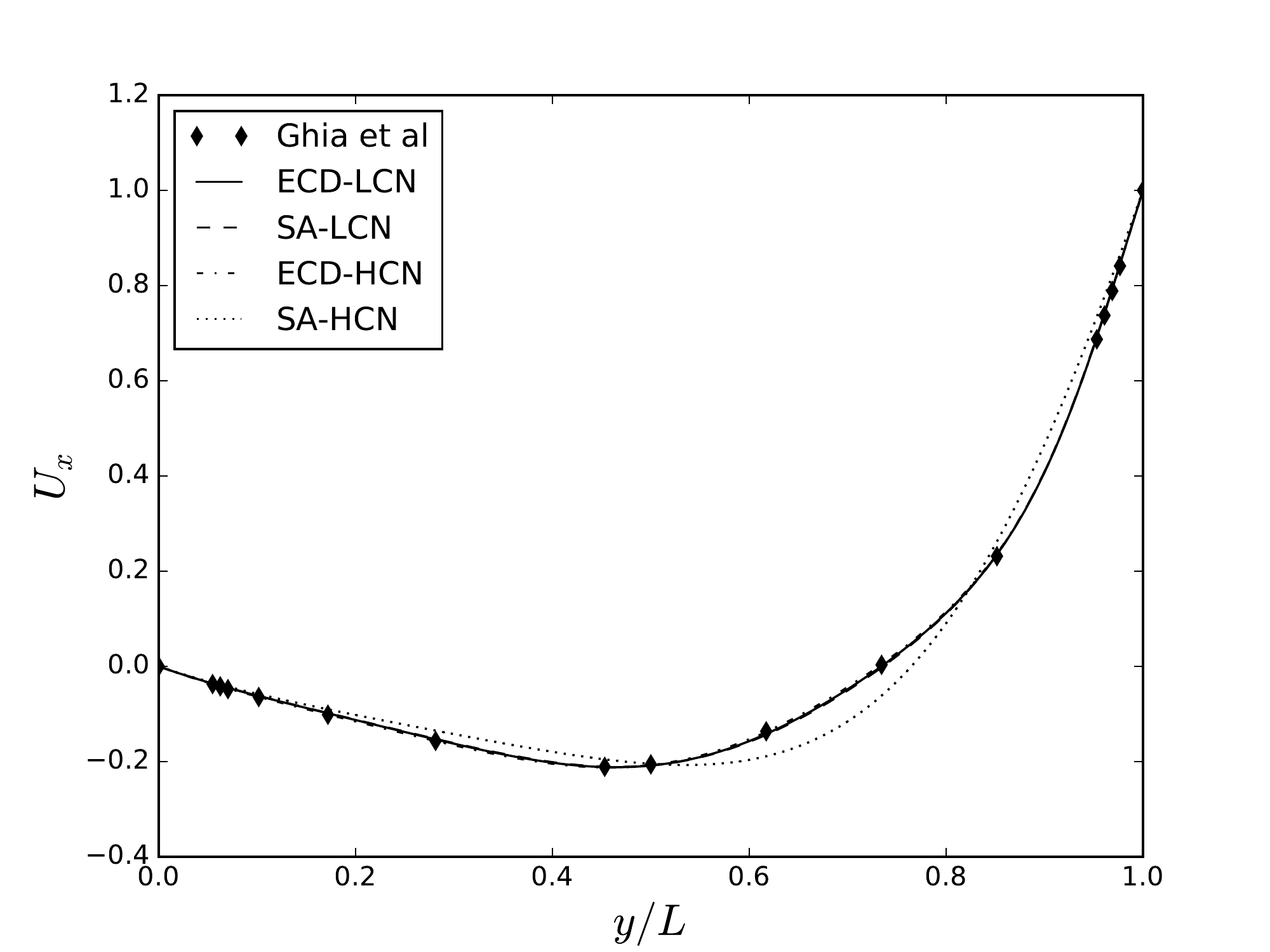}
	\end{subfigure}
	\begin{subfigure}[htbp]{0.4\textwidth}
		\caption{\label{fig:CavHln}Velocity profiles of $U_y$ along the horizontal line through cavity center}%
		\includegraphics[width=\textwidth]{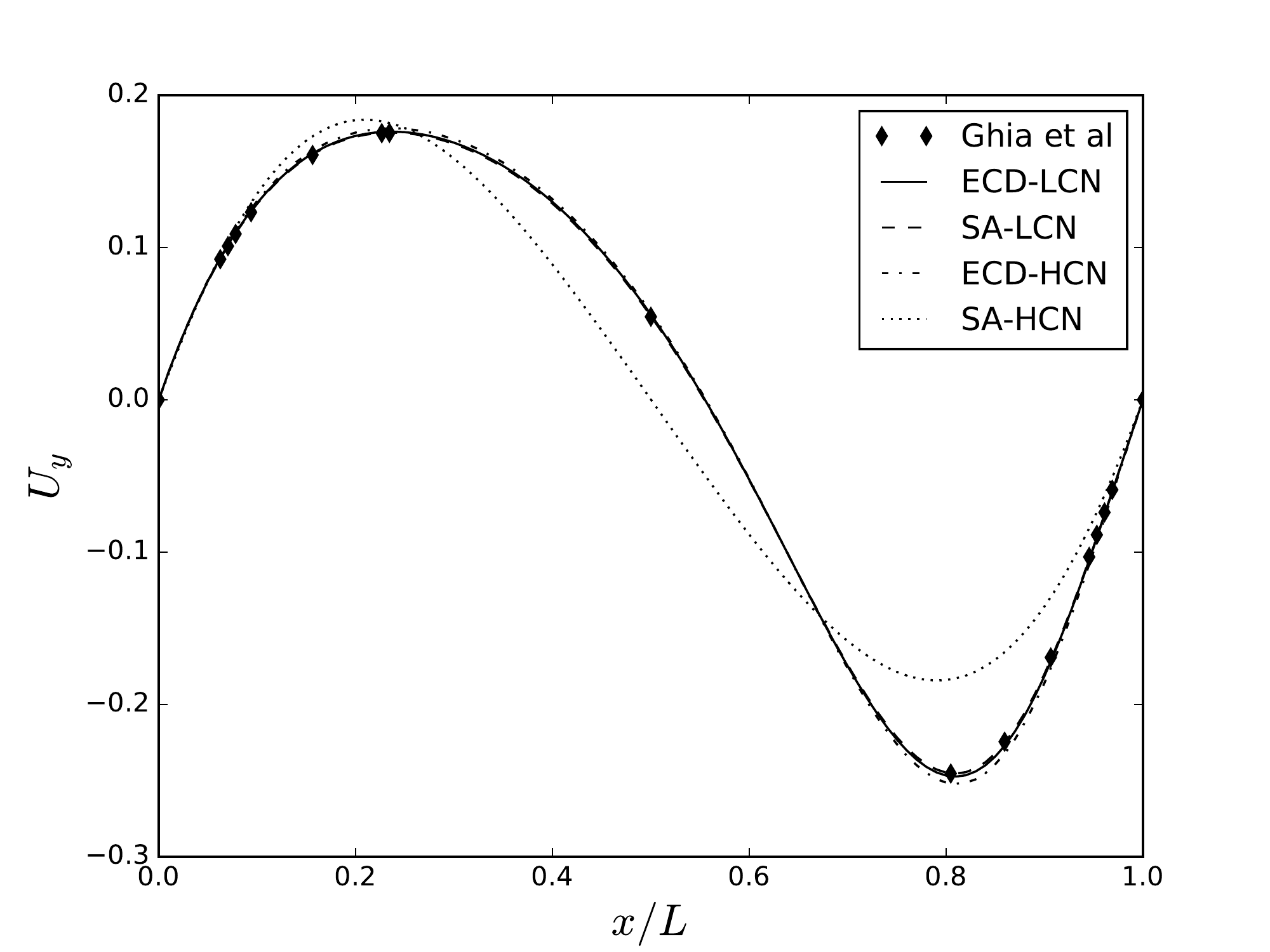}
	\end{subfigure}
	\caption{Velocity profiles along the lines through cavity center. The line with label {\it Ghia et al} is the calibration from \myRefInd{GhiaGhia-119} }\label{fig:cavLn}
\end{figure}

\begin{figure}[htbp]
	\centering
	\begin{subfigure}[htbp]{0.22\textwidth}
		\caption{ECD under LCN}\label{fig:CavL-ECD}
		\includegraphics[width=\textwidth]{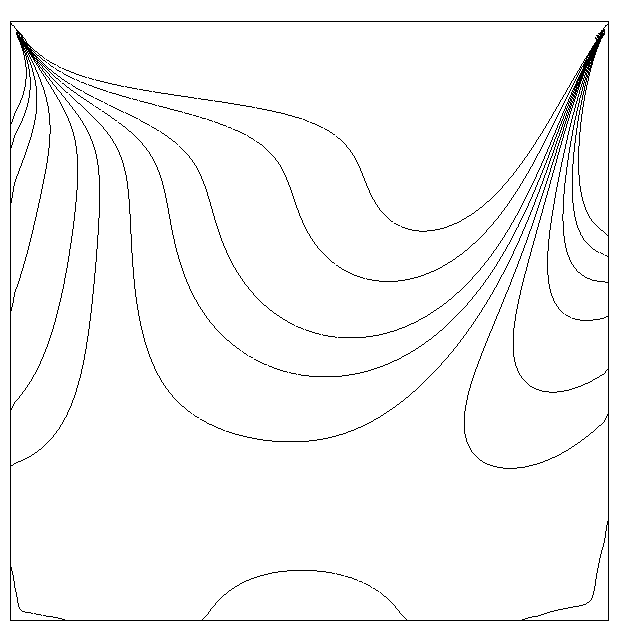}
	\end{subfigure}%
	\begin{subfigure}[htbp]{0.22\textwidth}
		\caption{SA under LCN}\label{fig:CavL-SA}
		\includegraphics[width=\textwidth]{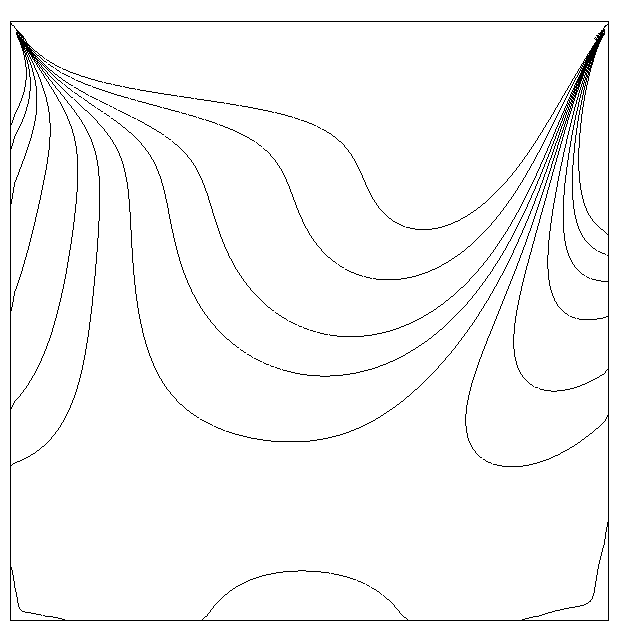}
	\end{subfigure}
	\begin{subfigure}[htbp]{0.22\textwidth}
		\caption{ECD under HCN}\label{fig:CavH-ECD}
		\includegraphics[width=\textwidth]{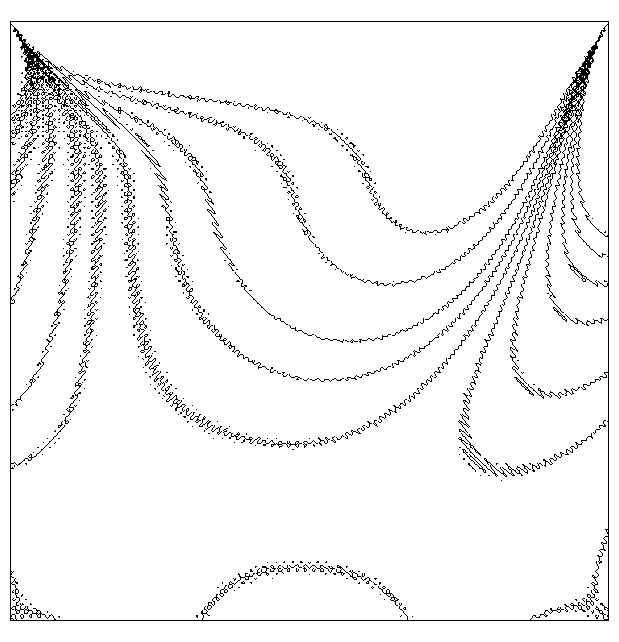}
	\end{subfigure}%
	\begin{subfigure}[htbp]{0.22\textwidth}
		\caption{SA under HCN}\label{fig:CavH-SA}
		\includegraphics[width=\textwidth]{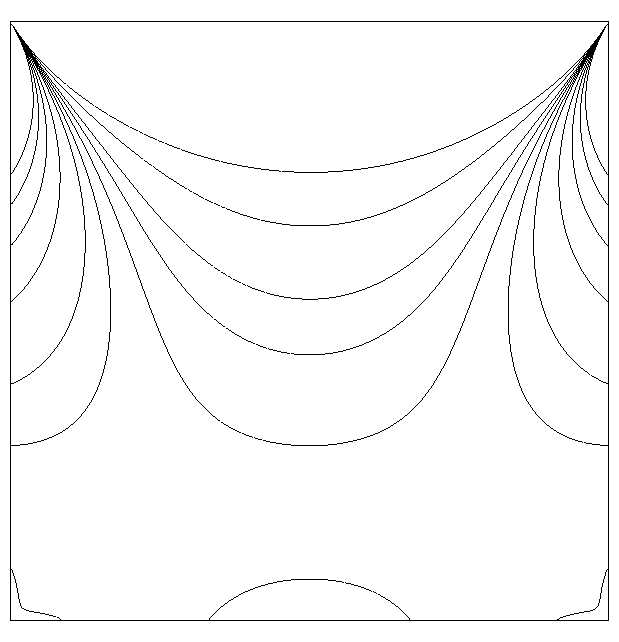}
	\end{subfigure}
	\caption{Vorticity contours of the cavity flow. The values of vorticity along contours are $\rm{0}$, $\pm \rm{0.5}$, $\pm \rm{1.0}$, $\pm \rm{2.0}$, $\pm \rm{3.0}$, $\rm{4.0}$, $\rm{5.0}$.}\label{fig:cavConVor}
\end{figure}

The simulation is configured as Table~\ref{tab:cav2d}. The cavity sides and corners employ the regularized boundary\mycite{LattChopard-124}. Fig.~\ref{fig:cavLn} illustrates the velocity profiles along the vertical and horizontal lines through cavity center.
The models keep consistent with their performance in \mySecRef{subsec:couet}, under \mrvs{LCN} , SA and ECD share a close profile; under \mrvs{HCN}, ECD is stick with the \mrvs{LCN} profile while SA fails. To give a full picture of the calculations, the vorticity contours are plotted in Fig.~\ref{fig:cavConVor}, which can be quantified by post-processing software compared to streamline. The values of vorticity along contours are $\rm{0}$, $\pm \rm{0.5}$, $\pm \rm{1.0}$, $\pm \rm{2.0}$, $\pm \rm{3.0}$, $\rm{4.0}$, $\rm{5.0}$. As Fig.~\ref{fig:cavConVor} illustrates, under LCN, SA (seeing Fig.~\ref{fig:CavL-SA}) and ECD (seeing Fig.~\ref{fig:CavL-ECD}) have the same vorticity contour; under HCN, though the contour of ECD in Fig.~\ref{fig:CavH-ECD} is a bit dispersed, the shape agrees well with LCN result while SA (seeing Fig.~\ref{fig:CavH-SA}) is quite different.


\begin{table}[htbp]
	\centering
	\caption{Configurations of lid-driven cavity flow. The cavity side length equals $1.0 \rm{m}$, the driven-lid velocity is $1.0 \rm{m/s}$}\label{tab:cav2d}
	\begin{tabular}{l ccccc}
		\hline\hline
		\noalign{\smallskip}
		CN\footnote{CN stands for collision number} & Grid &  & Kinematic & Collision \\[-1ex]
		type & number\footnote{Grid number is the mesh number on cavity side length $L$} & \raisebox{1.5ex}{Timestep} & viscosity & number \\
		\noalign{\smallskip}
		\hline
		% after \\: \hline or \cline{col1$-$col2} \cline{col3$-$col4} ...
		\noalign{\smallskip}
		LCN &  & 1.00E$-$3 &  & 5.09E$-$1 \\[-1ex]
		HCN & \raisebox{1.5ex}{257} & 1.00E$-$6 & \raisebox{1.5ex}{1.00E$-$2} & 5.09E$+$2 \\
		\noalign{\smallskip}
		\hline\hline
	\end{tabular}
\end{table}

\rvs{The results of SA model illuminate that, under LCN, as the collision change is minimal during a steptime, the constant equilibrium distribution assumption in SA LB equation can properly approximate its temporal evolution, meanwhile under HCN, due to the great collision change, it seriously deviates from the physical temporal evolution. 
The accurate result of ECD model under HCN indicates that the modification part in ECD model could accurately approximate the change of equilibrium distribution caused by intensive collision. The different numerical performances of SA and ECD LB equation under HCN confirm the effect of $g\left(t'\right)$ model on accuracy of LB equation.
Compared with unsteady Couette flow in \mySecRef{subsec:couet}, the governing equations of lid-driven cavity flow, i.e. full incompressible-flow Navier-Stokes equations, is more complicated and close to a real CFD problem. Thus the consistence of models' numerical performances between unsteady Couette and lid-driven cavity flow validates the universality of our assertion on the accuracy of SA and ECD LB equation under different collision numbers.  
Hence, we can conclude that, for a fluid calculation,   SA and ECD are both available under LCN meanwhile ECD is a better choice under HCN.}

\section{Conclusion}
\label{sec:con}
In this paper we propose \mrvs{a mathematically rigorous but physically general LB theory,  ACI LB theory.}  In ACI LB theory, we assume that BGK-Boltzmann equation is exact kinetic equation behind Navier-Stokes continuum and momentum equations and construct LB equation by analytically integrating BGK-Boltzmann equation along characteristics. Since the integral of $g\left(t'\right)$ along characteristics can not be analytically solved, the approximation is employed.
To \mrvs{ensure} that the evolution of LB equation does not depart the physical process too much, we restrict the value of  $g\left(t'\right)$ at $t'=0$ to $g^n$ so that the transient equation of particle interaction with respect to $\left( {\vec r,\vec \xi ,t} \right)$  is exact BGK-Boltzmann equation.
In ACI LB theory, the $g\left(t'\right)$ model is the determinant of LB equation accuracy and collision number is the characteristic parameter of equation, depicting the particle-interacting intensity in the time span. 
\mrvs{As a demonstration, we recover the approximating models behind popular LB equations, i.g. SA, Boesh-Karlin, ECD, DCD and He-Chen-Dong LB equation, and numerically analyze them,} where Boesh-Karlin and He-Chen-Dong equation are replaced by SA and ECD since they are equivalent respectively after removing the implicitness of $g^{n+1}$.  

To conclude, we highlight the following conclusions drawn from our work:
\begin{enumerate}[label={$\bullet$ }]  
	\item \rvs{LB equation can be constructed through analytically integrating BGK-Boltzmann equation along characteristics with approximated temporal evolution of equilibrium distribution $g\left(t'\right)$, avoiding the approximation and truncation in classical schemes\mycite{Dellar-2074, HeLuo-2016, SterlingChen-252, HeChen-173}.}
	\item \rvs{The kinetic theory depicted by BGK-Boltzmann equation is quite general, far beyond the discussion in Boesh-Karlin scheme\mycite{BoeschKarlin-2075}. By tuning $g\left(t'\right)$ model, most popular LB equations can be recovered as analytical characteristic integrals of BGK-Boltzmann equation, including the questioned LBGK equation.} 	
	\item \rvs{LB equation is identified by its $g\left(t'\right)$ model. The approximation accuracy of $g\left(t'\right)$ model determines LB equation accuracy. Among presented $g\left(t'\right)$ models, ECD is a better choice for numerical simulations.}
	\item The characteristic parameter of LB equation is collision number, $\Delta t/\lambda$, instead of the traditional relaxation time, $1/\tau$, which is merely a reflection of  $g\left(t'\right)$ model. It depicts the particle-interacting intensity in the time span of LB equation, independent from detail $g\left(t'\right)$ model. And the over relaxation time problem is merely a manifestation of ECD model under high collision number.	    
\end{enumerate}

\addcontentsline{toc}{chapter}{Acknowledgment}
\section*{Acknowledgment}
Huanfeng Ye would like to express his gratitude to Dr. Zecheng Gan for helpful discussion, and to Dr. Mathias J. Krause for generous support on OpenLB software. Huanfeng Ye is also grateful to the OpenLB team for their open source sharing (www.openlb.net).

\addcontentsline{toc}{chapter}{References}

%\bibliographystyle{unsrt}
%\bibliographystyle{cpb}
%%\bibliographystyle{ws-ijmpc}
%\bibliography{TLBRef.bib}

\end{document}